\documentclass{mn2e}

\usepackage{times}
\usepackage{amssymb}
\usepackage{epsfig}

\begin{document}
\hsize=6truein
          
\title[The stellar populations in $z = 7 - 9$ galaxies from HUDF12]
{The UV continua and inferred stellar populations of galaxies 
at ${\bmath {\bf z \simeq 7 - 9}}$ revealed by the Hubble Ultra Deep Field 2012 
campaign}

\author[J.S.~Dunlop, et al.]
{J.S. Dunlop$^{1}$\thanks{Email: jsd@roe.ac.uk}, A.B. Rogers$^{1}$, R.J. McLure$^{1}$,
R.S. Ellis$^2$, B.E. Robertson$^3$, A. Koekemoer$^{6}$, \and P. Dayal$^{1}$, 
E. Curtis-Lake$^1$, V. Wild$^{1,4}$, S. Charlot$^{5}$, 
R.A.A. Bowler$^{1}$, M.A. Schenker$^{2}$, \and
M. Ouchi$^{7}$, Y. Ono$^{7}$, M. Cirasuolo$^{1,8}$, 
S.R. Furlanetto$^{9}$, D.P. Stark$^{3}$, T.A. Targett$^1$,
\and E. Schneider$^{3}$.\\
\footnotesize\\
$^{1}$ SUPA\thanks{Scottish Universities Physics Alliance}, 
Institute for Astronomy, University of Edinburgh, 
Royal Observatory, Edinburgh, EH9 3HJ\\
$^{2}$ Department of Astrophysics, California Institute of Technology, MS 249-17, Pasadena, CA 91125, USA\\
$^{3}$ Department of Astronomy and Steward Observatory, University of Arizona, Tucson, AZ 85721, USA\\
$^{4}$ School of Physics and Astronomy, University of St Andrews, North Haugh, St Andrews, KY16 9SS\\ 
$^{5}$ UPMC-CNRS, UMR7095, Institute d'Astrophysique de Paris, F-75014, Paris, France\\
$^{6}$ Space Telescope Science Institute, Baltimore, MD 21218, USA\\
$^{7}$ Institute for Cosmic Ray Research, University of Tokyo, Kashiwa City, Chiba 277-8582, Japan\\
$^{8}$ UK Astronomy Technology Centre, Royal Observatory, Edinburgh, EH9 3HJ\\
$^{9}$ Department of Physics \& Astronomy, University of California, Los Angeles, CA 90095, USA}
\maketitle

\begin{abstract}
We use the new ultra-deep, near-infrared imaging of the Hubble Ultra-Deep Field 
(HUDF) provided by our UDF12 {\it Hubble Space Telescope} (HST) WFC3/IR 
campaign to explore the rest-frame ultraviolet 
(UV) properties of galaxies at redshifts 
$z > 6.5$. We present the first unbiased measurement of the
average UV power-law index, $\langle \beta \rangle$, 
($f_{\lambda}~\propto~\lambda^{\beta}$) for faint galaxies at $z \simeq 7$, 
the first meaningful measurements of 
$\langle \beta \rangle$ at $z \simeq 8$, and tentative 
estimates for a new sample of galaxies 
at $z \simeq 9$.
Utilising galaxy selection 
in the new F140W ($J_{140}$) imaging to minimize colour bias, and applying both  
colour and power-law estimators of $\beta$, 
we find $\langle \beta \rangle = -2.1 \pm 0.2$ at $z \simeq 7$ for galaxies 
with $M_{UV} \simeq -18$. This means 
that the faintest galaxies uncovered at this epoch have, 
{\it on average}, UV colours 
no more extreme than those displayed by the bluest star-forming galaxies at  
low redshift. At $z \simeq 8$ we find a similar value, $\langle \beta \rangle = -1.9 \pm 0.3$.
At $z \simeq 9$, we find $\langle \beta \rangle = -1.8 \pm 0.6$,
essentially unchanged from $z \simeq 6-7$ (albeit highly uncertain).
Finally, we show that there is as yet no evidence for a significant {\it intrinsic} 
scatter in $\beta$ within our new, robust $z \simeq 7$ 
galaxy sample. Our results are most easily explained by a population of steadily star-forming
galaxies with either $\simeq$ solar metallicity and zero dust, or moderately
sub-solar ($\simeq 10-20$\%) metallicity with modest dust obscuration ($A_V
\simeq 0.1-0.2$). This latter interpretation
is consistent with the predictions of a state-of-the-art galaxy-formation simulation,
which also suggests that a significant population
of very-low metallicity, dust-free galaxies with $\beta \simeq -2.5$ may not emerge 
until $M_{UV} > -16$, a regime likely to remain inaccessible until the {\it James Webb Space Telescope}.

\end{abstract}

\begin{keywords}
galaxies: high-redshift - galaxies: evolution - galaxies: formation - 
galaxies: stellar populations - cosmology: reionization
\end{keywords}

\section{INTRODUCTION}

The revolution in very-deep, near-infrared imaging provided by the 2009 refurbishment of 
the {\it Hubble Space Telescope} (HST) with the Wide Field Camera 3 (WFC3/IR) 
has enabled the discovery and study of the first substantial 
samples of galaxies at $z > 6.5$ (see, e.g., Dunlop 2012 for a review). Following the  
instant success of the initial deep $Y_{105}, J_{125}, H_{160}$ 
UDF09 imaging (GO 11563; PI: G. Illingworth) of the 
Hubble Ultra-Deep Field (HUDF; Beckwith et al. 2006) and associated parallel fields 
(e.g. Oesch et al. 2010; Bouwens et al. 2010a, 2011; McLure et al. 2010, 2011; Finkelstein et al.
2010; Bunker et al. 2010), WFC3/IR has been used to conduct wider-area surveys for more luminous galaxies at $z = 7-8$, 
both through the CANDELS Treasury programme (Grogin et al. 2011; Koekemoer et al. 2011; Grazian et al. 2012; Oesch et al. 
2012), and through parallel imaging programmes such 
as the BoRG survey (e.g. Bradley et al. 2012). Most recently, attention 
has been refocussed on pushing to even fainter magnitudes 
and still higher redshifts, either with the assistance of gravitational 
lensing (e.g. through the CLASH Treasury Programme; Zheng et al. 2012; Coe et al. 2013),
or through our own ultra-deep WFC3/IR imaging in the HUDF 
(GO 12498; PI: R. Ellis, hereafter UDF12).

Our recently-completed 128-orbit UDF12 observations 
reach 5-$\sigma$ detection limits 
of $Y_{105}$ = 30.0, $J_{125}$ = 29.5, $J_{140}$ = 29.5, $H_{160}$ = 29.5 (after combination with the UDF09 data), and are the deepest 
near-infrared images ever taken (Ellis et al. 2013). A detailed description
of the UDF12 data-set is provided by Koekemoer
et al. (2013), and the final reduced images will be available
on the team web-page\footnote{http://udf12.arizona.edu}.

These new, ultra-deep, multi-band near-infrared images 
have already yielded the first significant sample of galaxies at $z \simeq 9$, including a possible 
candidate at $z \simeq 12$ (Ellis et al. 2013). The discovery of galaxies at $z > 8.5$ was a key design 
goal of this programme, and motivated the first inclusion of deep $J_{140}$ imaging in the HUDF (elsewhere, 
$J_{140}$ imaging 
has also played a key role in enabling CLASH to yield convincing galaxy candidates out to $z \simeq 10.7$;
Coe et al. 2013). However, the additional deeper $H_{160}$ imaging, and the ultra-deep $Y_{105}$ 
imaging has also been crucial in enabling the more robust selection of objects at $z \simeq 7$ and $z \simeq 8$
(with improved photometric redshifts; McLure et al. 2013), 
and a push to still fainter magnitudes. Another key goal of the UDF12 programme 
was therefore to use the resulting improved samples, coupled with the more accurate 4-band near-infrared photometry, to 
undertake a new and unbiased study of the rest-frame ultraviolet (UV) spectral energy distributions (SEDs) of 
faint galaxies at $z > 6.5$.

This paper is thus focussed on revisiting the study of the rest-frame UV SEDs of galaxies, and in particular 
their UV continuum slopes,  $\beta$ (where $f_{\lambda} \propto \lambda^{\beta}$; e.g. Calzetti, 
Kinney \& Storchi-Bergmann 1994; 
Meurer et al. 1999), 
armed with the best-available, 
near-infrared data required for this measurement at $z \simeq 7, 8$ (and, for the first time, at $z \simeq 9$). 
Because the objects uncovered by HST in the HUDF at these redshifts are too faint for informative near-infrared 
spectroscopy with current facilities, a broad-band determination of the UV continuum slope $\beta$ at present 
offers the only practical way of gaining insight into the rest-frame UV properties of the early populations 
of galaxies emerging in the young Universe. This (in principle simple, but in practice tricky) measurement 
is of astrophysical interest for a number of reasons. 

First, certainly at more modest redshifts, $\beta$ has been shown to be a good tracer of dust extinction 
in galaxies, as it has been demonstrated to be well correlated with excess far-infrared {\it emission} from 
dust (e.g. Meurer et al. 1999; Reddy et al. 2012; Heinis et al. 2013). The reason this works 
is that (as we again demonstrate later in this paper) an actively star-forming galaxy of $\simeq $ solar metallicity would be expected to 
display $\beta \simeq -2$ ($\equiv$ zero colour in the AB magnitude system) in the absence of dust, and so any significant deviation 
to redder values can be viewed as a signature of significant dust extinction (albeit the relation between far-infrared 
dust-emission and UV-derived dust emission is not expected to be perfect, given that different regions 
of the galaxies might be observed in such widely-separated wavelength regimes; e.g. Wilkins et al. 2012; Gonzalez-Perez et al. 2013). 
These results have been used to interpret
the apparent steady progress towards lower average values of $\langle \beta \rangle$ with increasing 
redshift (from $z \simeq 4$ to $z \simeq 7$) in terms of monotically-decreasing average dust extinction, 
with important implications for the inferred cosmological evolution of star-formation activity (e.g. Hathi et al. 2008; 
Bouwens et al. 2009, 2012; Castellano et al. 2012; Finkelstein et al. 2012). 

Second, $\beta$ is obviously also a function of age (see, for example, Fig. 2 in Rogers, Dunlop 
\& McLure 2013), although (as we explicitly demonstrate later in this 
paper) the sensitivity is not very strong 
for young, quasi-continuously star-forming sources, and very 
blue values of $\beta$ can certainly only be achieved for very young stellar populations.  

Third, $\beta$ can be used as an indicator of metallicity. In practice of course the impact of metallicity 
and dust extinction can be degenerate for redder values of $\beta$, but blue values significantly lower
than $\beta \simeq -2$ are an indicator of a low-metallicity stellar population. For example
NGC1703, one of the bluest local star-forming galaxies with $\beta \simeq -2.3$, is generally intepreted
as being dust-free, with a significantly sub-solar metallicity (Calzetti et al. 1994), as is 
the low-mass galaxy BX418 at $z \simeq 2.3$ for which Erb et al. (2010) report $\beta = -2.1$, $E(B-V) \simeq 0.02$,
and $Z \simeq 1/6\,{\rm Z_{\odot}}$.    

Fourth, $\beta$ is influenced by the extent to which the emission from the combined photospheres of the 
stars in a galaxy is `contaminated' by nebular continuum. Nebular continuum emission is significantly 
redder than the star-light from a very young, low-metallicity stellar population
(e.g. Leitherer \& Heckman 1995), and so,  
given other information (e.g. Stark et al. 2013; Labb\'{e} et al. 2013) or assumptions, $\beta$ can in principle be used to estimate
(or correct for)  the level of 
nebular emission in a young galaxy. This in turn can set constraints on the inferred escape-fraction 
($f_{esc}$) of Hydrogen-ionizing photons. It is the rate-density of such photons that requires to be 
estimated to chart the expected progress of cosmic reionization by young galaxies (e.g. Robertson et al. 2010), 
but such photons are not directly observable during the epoch of reionization. 

Thus, as discussed by many authors, there is a strong motivation 
for attempting to measure $\beta$ as accurately as possible, but the 
interpretation of the results can clearly be problematic given the degeneracies involved. 
Interestingly, however, the degree of complication 
in interpretation is result-dependent. In particular, as highlighted 
by Schaerer (2002) and Bouwens et al. (2010b), the discovery of {\it extremely} blue values of $\beta \simeq -3$ 
would offer a fairly clean and powerful result, because such values can only be produced by a stellar population 
which is simultaneously very young, of extremely low metallicity, dust-free, and also free of significant  
nebular emission (corresponding to a very high escape fraction for ionizing photons). Since these are exactly the 
combined properties which might be expected of the first galaxies (which possibly commenced the reionization of 
the Universe; e.g. Paardekooper, Khochfar \& Dalla Vecchia 2013; 
Mitra, Ferrara \& Choudhury 2013) the measurement of $\beta$ during the first billion years of cosmic time has been a key focus
of several recent studies of galaxies at $z \simeq 7$ (Bouwens et al. 2010b, 2012; Dunlop et al. 2012; 
Finkelstein et al. 2010, 2012; McLure et al. 2011; Wilkins et al. 2011; Rogers, Dunlop \& McLure 2013). 

However, as discussed in detail by Dunlop et al. (2012) and Rogers
et al. (2013), previous attempts to determine the UV spectral slope 
at faint magnitudes ($M_{UV} > -19$) have inevitably been afflicted  
by bias. The interested reader is referred to these papers for 
detailed discussions and simulations, but there are three key points 
to consider, 
each related to photometric scatter and the resulting impact on 
derived {\it average} values ($\langle \beta \rangle$).

First, the selection band can bias the result if source selection 
is pushed to the 5-$\sigma$ limit, and the primary selection 
band is then also involved in colour determination. 
For example, imposition of 
a $J_{125}$ flux density threshold, as applied by Bouwens et al. (2010b),
inevitably leads to a blue-bias in the derived average value of 
$\langle \beta \rangle$ if $\beta$ is based on $J_{125}-H_{160}$ colour.

Second, the classification of objects as {\it robust} high-redshift Lyman-break
galaxies can also yield a subtle bias towards bluer values of $\beta$. 
Again this is only really an issue for the derivation of average values 
from individual measurements with substantial photometric scatter. The point 
is that, while both colour-colour selection (e.g. Bouwens et al. 2010a, 2011)
and photometric redshift selection (e.g. McLure et al. 2010, 2011) are 
sufficiently inclusive to include virtually all plausible star-forming
galaxy SEDs at high redshift, photometric scatter can lead to some genuine 
high-redshift galaxies being misclassified as being at much lower redshift.
Because this only happens when the scatter yields erroneously red colours, the  
result can be clipping of the red end of the {\it observed} colour 
distribution, yielding a blue bias in the derived average $\langle \beta
\rangle$. Rogers et al. (2013) 
show that this bias is, unsurprisingly,
essentially identical for colour selection and photometric-redshift selection
provided {\it all} galaxies with a plausible high-redshift photometric solution are
retained in the latter approach. If attention
is confined to the most robust photometrically-selected high-redshift
galaxies, then the bias is unfortunately 
inevitably more extreme (because a very blue
colour longward of the putative Lyman break essentially guarantees a
robust high-redshift solution; see Dunlop et al. 2012).

Third, if, as advocated by Finkelstein et al. (2012), 
$\beta$ is derived from galaxy spectral energy distribution (SED) models (e.g. Bruzual \& Charlot 
2003; hereafter BC03), the result can be a red bias in $\langle \beta \rangle$. At first sight, the use 
of galaxy SEDs seems sensible, but the problem is that model galaxy SEDs never 
produce $\beta$ significantly bluer than $\beta \simeq -3$. While there 
is good reason to believe that real galaxy SEDs can never actually yield
$\beta$ significantly bluer than $\beta \simeq -3$, if one wants to 
compute a population average from a photometrically-scattered set of
objects, it is (as already emphasized) 
important not to artificially clip one end of the {\it observed} distribution.
In this case the effect of insisting on fitting plausible galaxy SEDs is to clip the 
blue end of the distribution, because any object which displays, say, $\beta
= -5$ will be corrected back to $\beta = -3$ (or whatever the most extreme
$\beta$ contained in the galaxy SED library happens to be). The result is a 
red bias in the average $\langle \beta \rangle$. Thus, as 
explicitly demonstrated by Rogers et al. (2013), 
the most robust way to determine an unbiased value of 
$\langle \beta \rangle$ is via a pure power-law fit to the appropriate
photometry.

The primary aim of this paper is not to revisit these bias issues but, 
rather, to use the new, deep, multi-band near-infrared photometry 
provided by the UDF12 WFC3/IR imaging campaign to avoid them, 
and deliver the first straightforward, unbiased measurement of 
$\langle \beta \rangle$ for faint galaxies at $z \simeq 7$ 
(comparable to, but somewhat fainter than the 
luminosity regime where $\langle \beta \rangle \simeq -3$ 
was originally claimed by Bouwens
et al. 2010b). The UDF12 campaign was, in part, designed with this goal
in mind. First, the increase in depth and the addition of 
an extra passband $J_{140}$ allows significantly more 
accurate measurements of UV slope down to $M_{UV}
\simeq -17$. Second, the introduction of the deep 
$J_{140}$ imaging allows object selection to be based primarily on a band 
which has minimal influence on derived UV slope (whether derived by 
$J_{125} - H_{160}$ colour, or by power-law fitting through 
$J_{125}$, $J_{140}$, $H_{160}$). 

A second aim of this paper is then to exploit both the new UDF12 
photometry, and the new $z \simeq 8$ and $z \simeq 9$
galaxies uncovered by McLure et al. (2013) and
Schenker et al. (2013) in our UDF12 programme 
(Ellis et al. 2013; Koekemoer et
al. 2013), to present the first meaningful measurements of 
$\langle \beta \rangle$ at $z \simeq 8$ and $z \simeq 9$. Inevitably 
these first results on $\beta$ at even higher redshift apply to 
somewhat brighter absolute magnitudes ($M_{UV} \simeq -18$) than the faintest 
bin explored at $z \simeq 7$, but nevertheless it is of interest to explore
the behaviour of $\beta$ back to within $\simeq 0.5$\,Gyr of the big bang.
Measurements of $\beta$ at even earlier epochs will not be possible
until the launch of the {\it James Webb Space Telescope} (JWST) 
and the advent of ground-based Extremely Large Telescopes (ELTs).

Finally, a third aim of this paper is to attempt to move beyond 
the determination of $\langle \beta \rangle$ at $z \simeq 7$, and explore 
whether the improved accuracy of individual measurements of $\beta$ afforded
by the UDF12 data provide any evidence for a significant intrinsic 
scatter in $\beta$ at $z \simeq 7$. The impact of UDF12 on 
the evidence for a colour-magnitude
relation, and its potential evolution over a broader redshift range 
$z \simeq 4$ to $z \simeq 7$ will be considered in a separate paper 
(Rogers et al., in preparation).

This paper is structured as follows. In Section 2 we briefly summarize 
the new UDF12+UDF09 dataset, and the way in which our new 
high-redshift galaxy samples spanning the redshift range 
$6.5 < z < 12$ have been selected. Then, in Section 3 we present 
straightforward (but now essentially unbiased) 
`traditional' colour measurements of $\beta$ at $z \simeq 7$ (to 
aid comparison with previous studies) and also for the first time at 
$z \simeq 8$ and $z \simeq 9$. In Section 4 we then present our `best' 
measurements of $\beta$ based on the multi-band power-law fitting 
as developed and advocated in Rogers et al. 
(2013). Here we also draw on the results 
of our simulations to demonstrate the absence of any substantial  
bias in our measurements, and to correct for any minor residual effects.
We consider the astrophysical implications of our results in Section 5, 
including a comparison with the predictions of the latest hydrodynamical
models of galaxy evolution. Finally our conclusions are summarized in Section
6. All magnitudes are quoted in the AB system (Oke 1974) and any
cosmological calculations assume $\Omega_M = 0.3$, $\Omega_{\Lambda} = 
0.7$, and $H_0 = 70\,{\rm km s^{-1} Mpc^{-1}}$.

\section{Galaxy Samples}

\subsection{UDF12 high-redshift galaxy sample selection}

New HUDF galaxy samples were selected from the final UDF12+UDF09 dataset 
as follows. First, SExtractor (Bertin \& Arnouts 1996) 
was used to select any source which yielded 
a $>5$-$\sigma$ detection in any one of the final single-band WFC3/IR $Y_{105}$, $J_{125}$, 
$J_{140}$ or $H_{160}$ images, or in any contiguous stacked combination of them 
(i.e. $Y_{105}$+$J_{125}$, $J_{125}$+$J_{140}$, $J_{140}$+$H_{160}$, 
$Y_{105}$+$J_{125}$+$J_{140}$,   
$J_{125}$+$J_{140}$+$H_{160}$,
$Y_{105}$+$J_{125}$+$J_{140}$+$H_{160}$). The catalogues from these 
ten alternative detection runs were next merged to form a parent sample which 
was then culled by rejecting any source which showed a $>2$-$\sigma$ detection in any of 
the three shortest-wavelength ($B_{435}$, $V_{606}$, $i_{775}$) 
deep HST ACS optical images of the HUDF. This process means that the resulting galaxy
sample should remain complete beyond $z = 6.4$ (although it will also yield 
some galaxies in the redshift range $z \simeq 6 - 6.4$; see McLure et al. 2011, 2013).

Multi-band aperture photometry was then performed at the position of each object (as 
determined from the detection image which yielded the highest signal:noise ratio detection),
using `matched' circular apertures designed to contain 70\% of the flux density from a
point source. The appropriate aperture diameters as used are 0.5\,arcsec at 
$H_{160}$, 0.47\,arcsec at $J_{140}$, 0.44\,arcsec at $J_{125}$, 0.40\,arcsec at $Y_{105}$,
and 0.3\,arcsec for the ACS $z_{850}$ photometry. The WFC3/IR and $z_{850}$ 
photometry was then all aperture-corrected 
to 82\% of total to correctly match the 
photometric limits achieved for point sources 
in the bluer ACS bands within a 0.3-arcsec diameter aperture.
In addition, 
{\it Spitzer} IRAC photometry 
at 3.6 and 4.5\,$\mu$m was included for each source. This was based on a new deconfusion 
analysis of the deepest available IRAC HUDF imaging (Labb\'{e} et al. 2012) using the 
method described by McLure et al. (2011), and the final UDF12 $H_{160}$ 
image as the best available template for the galaxies to be fitted to the sky as seen in the 
highly-confused IRAC imaging. In practice, for all but a few sources, this yielded 
formal non-detections.

Photometric redshifts (with associated probability 
distributions) were then derived on the basis of this 10-band photometry, using the 
method described in McLure et al. (2011). Specifically, a wide range of galaxy SEDs 
from the BC03 models were utilised, limited only by insisting 
that galaxy ages were younger than the age of the Universe 
at each redshift) and dust extinction 
was allowed to float as high as $A_V = 4$. All flux-density measurements 
were utilised in this model fitting, even in bands where sources 
were undetected (including negative flux-density where measured,
thus ensuring a consistent derivation of $\chi^2$).

On the basis of the SED fitting, the sample was further refined 
by retaining only those sources which displayed a statistically-acceptable 
solution at $z > 6.4$ (i.e. redshift solutions 
with a formally acceptable value of $\chi^2$, in practice $\chi^2 < 15$).
At this stage all remaining candidates were visually inspected, and rejected from the 
catalogue if they lay too near the perimeter of the imaging, or too close to 
bright sources for reliable photometry (a cull that is reflected in the 
effective survey areas utilised by McLure et al. 2013 in the luminosity 
function analysis presented therein). A final visual check 
was also performed to remove any object which yielded 
a significant detection in a smoothed, stacked $B_{435}$+$V_{606}$+$i_{775}$
pseudo broad-band optical ACS image (in order to further minimize 
the number of low-redshift galaxy contaminants).

Finally, as in Dunlop et al. (2012), all of the 
ACS+WFC3/IR+IRAC SED fits were inspected, and the sources classified as 
ROBUST or UNCLEAR depending on whether the secondary, low-redshift solution 
could be excluded at $>$2-$\sigma$ significance, as judged by 
$\Delta \chi^2 > 4$ between the secondary and primary redshift solution.
The final result of this process is a sample of 146 sources, of which 97
are labelled as ROBUST and 49 are UNCLEAR.
Absolute rest-frame UV magnitudes at $\lambda_{rest} \simeq 1500$\,\AA ($M_{1500}$)
have been 
calculated for all objects by integrating the spectral energy distribution
of the best-fitting evolutionary synthesis model 
through a synthetic `narrow-band' filter of rest-frame width 100\,\AA\
(see McLure et al. 2011).

\subsection{Further sample refinement for ${\bf \beta}$ analysis}
 
For the specific purpose of the UV continuum slope analysis presented 
in this paper, the sample was then split into three redshift bins, 
yielding 116 galaxies at $z \simeq 7$ ($6.4 < z < 7.5$),  
24 galaxies at $z \simeq 8$ ($7.5 < z < 8.5$) and 6 galaxies at $z \simeq 9$
($8.5 < z < 10$).
Finally, 
to minimize any bias in $\beta$ introduced at the galaxy-selection stage,
we decided to limit the final galaxy catalogues at $z \simeq 7$ and $z \simeq 8$ to those
objects which yielded a $>$5-$\sigma$ detection in the $J_{140}$ band alone 
(i.e. $J_{140} < 29.5$ in the 0.47-arcsec diameter aperture). This reduced the number
of galaxies in each sub-sample to 45 at $z \simeq 7$ and 12 at $z \simeq 8$, but still allows 
us to probe galaxy luminosities down to $M_{UV} \simeq -17$ at $z \simeq 7$. This also means 
that we are here exploiting the extra depth of the additional UDF12 $Y_{105}$ and $H_{160}$ 
imaging primarily for the determination of more accurate measurements of $\beta$, rather 
than to push the detection threshold to the absolute limit. This somewhat conservative approach 
has the beneficial side-effect of reducing the fraction of UNCLEAR galaxies in the final sample to 
only $\simeq 10$\%, minimizing the need to consider the differences between analyses based on total 
or ROBUST-only samples.

All derived numbers, plots, and simulations presented hereafter assume the application
of this $J_{140}$ threshold at $z \simeq 7$ and $z \simeq 8$. However, at $z \simeq 9$,
application of this
threshold leaves only 2 objects, and so we do not apply it. In any case,  as 
outlined in Section 3.2, by $z \simeq 9$, $J_{140}$ 
is the bluest band involved in the $\beta$ estimation, and so its usefulness as an 
unbiased band for galaxy selection disappears. For this and other obvious reasons, the results 
we present on $\langle \beta \rangle$ at $z \simeq 9$ are treated separately, 
and should be regarded only as tentative/indicative
(as compared to the statistically-robust results we now provide 
at $z \simeq 7$ and $z \simeq 8$).

\begin{figure}
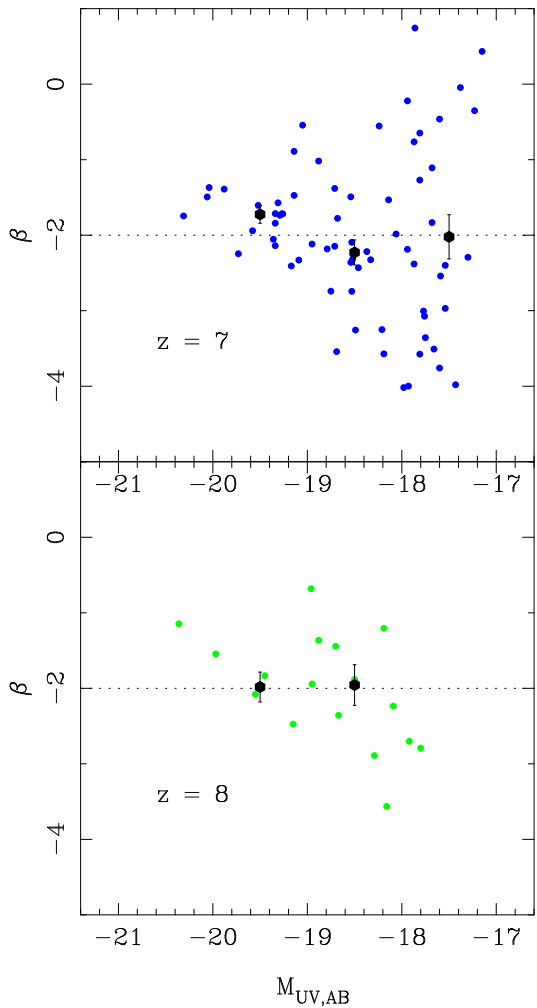

\centerline{\epsfig{file=fig1a.eps,width=7.0cm,angle=0}}
\vspace*{-0.47cm}
\centerline{\epsfig{file=fig1b.eps,width=7.0cm,angle=0}}
\caption{Individual measurements of UV continuum slope, $\beta$, 
at $z \simeq 7$ (upper panel, blue points) and
at $z \simeq 8$ (lower panel, green points) for the galaxies in the 
new UDF12 samples (as detailed in Section 2.2) plotted 
versus their UV absolute magnitudes
($M_{UV, AB} \equiv M_{1500}$).
The values of $\beta$ shown here are derived from the UDF12 
data using $J_{125}-H_{160}$ colours as described in Section 3.
The average values, along with standard errors
in the mean, are plotted (in black) for each 1-magnitude wide 
luminosity bin which contains $> 5 $ sources (see also Table 1). 
The UDF12 galaxy samples used have been 
confined to those objects which are detected at 
$> 5\sigma$ in the $J_{140}$-band, in order to minimize colour bias in 
the selection process ($J_{140}$ photometry is not used here in the 
determination of $\beta$). To help provide dynamic range, the 
samples at $M_{UV,AB} < -19$ have been supplemented with $>8$-$\sigma$ 
objects from the UDF09P1 and UDF09P2 parallel WFC3/ACS fields. Errors
for individual $\beta$ measurements are not shown, simply because the
typical error can be judged directly from the scatter in the plot (which,
it transpires, is effectively all due to photometric error; see Section 4.3). The 
$\beta$ values for the individual objects are provided in Table A1.}
\end{figure}

\begin{figure}
\centerline{\epsfig{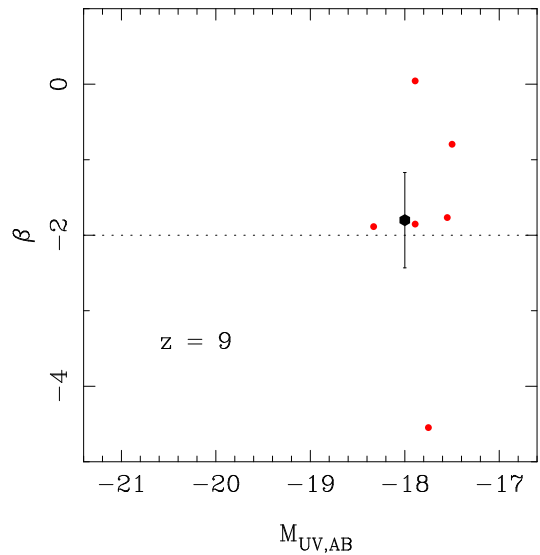}}
\caption{Individual measurements of UV continuum slope, $\beta$, (red points) 
as derived from our UDF12 
data using $J_{140}-H_{160}$ colour for the 6 galaxies at 
$z \simeq 9$ ($8.5 < z < 10$; Ellis et al. 2013), plotted against 
their UV absolute magnitudes
($M_{UV, AB} \equiv M_{1500}$).
The average value, $\langle \beta \rangle$,  at $M_{UV} \simeq -18$
is indicated by the black point, with the error-bar corresponding to 
the standard error in the mean (see Table 1).
This is the first attempt at such a measurement 
at this redshift, and the $J_{140}-H_{160}$ 
colour does not span a very large wavelength baseline. Moreover, with such 
a small sample at $z \simeq 9$, the statistical average is clearly 
not very robust. Nevertheless, the available information 
suggests that $\langle \beta \rangle$ at $M_{UV,AB} = -18$ at
$z \simeq 9$ has not changed dramatically from $z \simeq 7$, 
and is still consistent with $\beta= -2$.}
\end{figure}

\section{Two-band measurement of UV slopes}

The standard convention is to characterise the rest-frame UV 
continuum slope via a power-law index, $\beta$, where $f_{\lambda} \propto
\lambda^{\beta}$. For objects at $z \simeq 7$, the $Y_{105}$-band photometry 
could,
in principle, be contaminated by Lyman-$\alpha$ emission-line flux, and 
so it has been common practice to limit the measurement of $\beta$ 
at $z > 6.5$ to an estimator based on $J_{125} - H_{160}$ colour
(e.g. Bouwens et al. 2010b; Dunlop et al. 2012). Now, with the addition
of the new $J_{140}$ imaging from the UDF12 campaign,  
a three-band ($J_{125}$, $J_{140}$, $H_{160}$) power-law fit can be performed 
at $z \simeq 7$. As quantified in Rogers et al.  
(2013), a three-band power-law
fit is in fact the optimal way to determine $\beta$ for galaxies in this redshift regime,
and so we apply this method for the first time at these redshifts in Section 4.
However, to facilitate comparison with previous work, and 
to see the direct impact of our deeper photometry and 
our $J_{140}$ galaxy selection on the results, 
we first perform the standard $J_{125} - H_{160}$ colour estimation of $\beta$ 
on our new UDF12 $J_{140}$ filtered sample. Moreover, at $z \simeq 9$, 
only $J_{140}$ and $H_{160}$ are capable of sampling the continuum longward of
$\lambda_{rest} \simeq 1215$\,\AA, and so $\beta$ has to be based 
upon $J_{140} - H_{160}$ colour, 
rendering power-law fitting once again essentially
redundant.

The effective wavelengths of the filters of interest in this study
are $J_{125}$:$\lambda_{\rm eff}$=$12486$\AA, 
$J_{140}$:$\lambda_{\rm eff}$=$13923$\AA, and $H_{160}$:$\lambda_{\rm eff}$=$15369$\AA.
We note that these are the `pivot' wavelengths, incorporating
not only the filter transmission profiles, but also the full throughput 
of WFC3/IR including detector sensitivity as a function of wavelength.
There are of course a number of definitions of `effective wavelength'
for broad-band filters, 
but the `pivot' wavelength is the appropriate one for the present study, 
and in any case agrees with the alternative `mean' (source independent) 
wavelength of the filter to within better than 1\% (see, for example,
Tokunaga \& Vacca 2005). 

Adopting the above effective wavelengths, the appropriate equations for 
converting from near-infrared colour to $\beta$ are simply

\begin{equation}
\beta = 4.43(J_{125}-H_{160})-2
\end{equation}

\noindent
for measurements at $z \simeq 7 - 8$ 
based on $J_{125}-H_{160}$ colour, and 

\begin{equation}
\beta = 9.32(J_{140}-H_{160})-2
\end{equation} 

\noindent
for measurements at $z > 8.5$ which have to be based purely 
on $J_{140}-H_{160}$.

As already noted by Dunlop et al. (2012), 
equation (1) differs very slightly from the 
relation adopted by Bouwens et al. (2010b), which is  
$\beta = 4.29(J_{125}-H_{160})-2$ (presumably due to the adoption of slightly
different effective wavelengths). However, 
the differences in derived values of $\beta$ are completely insignificant
in the current context (e.g. for $J_{125}-H_{160} = -0.2$, the Bouwens et al. relation yields 
$\beta = -2.86$, while equation (1) yields $\beta = -2.89$).

Finally, we caution that equation (2) must be regarded with 
some scepticism. First, the relatively short wavelength-baseline 
provided by $J_{140}-H_{160}$ colour is reflected in the rather 
large coefficient by which colour 
(and hence also uncertainties in colour) must be multiplied
to yield an estimate of $\beta$. Second, whereas $J_{125}$ and $H_{160}$ 
provide independent samples of a galaxy SED, the $J_{140}$ and $H_{160}$ 
bands overlap, and hence the resulting measurements are inevitably 
correlated to some extent. For both these reasons equation (1) should 
be utilised rather than equation (2) whenever possible. Nevertheless,
out of curiosity, in Section 3.2 below  we apply equation (2) to the 6 objects in the 
$8.5 < z < 10$ sample to obtain a first direct observational estimate 
of $\langle \beta \rangle$ in this previously unexplored redshift regime.
\begin{table}
 \begin{center}

\caption{Derived average UV continuum slopes, $\langle \beta \rangle$, 
and standard errors as a function of absolute UV magnitude (in bins with 
$\Delta M_{UV} = 1$\, mag) and redshift.
The values given in column two are derived from simple two-band colours
as described in Section 3, and are only tabulated 
for luminosity bins that contain $>5$ galaxies from the 
UDF12 and/or UDF09 Parallel Fields (see Figs 1 and 2). 
The values given in columns three and four are derived
using the power-law fitting technique described in Section 4, and 
have been corrected for small residual biases as evaluated from the 
results of the end-to-end source injection, retrieval, 
and measurement simulations detailed in Section 4.1 (see Figs 3 and 4).
The galaxy samples analysed at $z \simeq 7$ and $z \simeq 8$ were restricted 
to objects detected at $> 5$-$\sigma$ in $J_{140}$ to minimize colour-selection
bias, as described in Section 2.2. The galaxy sample at $z \simeq 9$ was not 
restricted in this way (only 2 out of the 6 objects would remain). 
We also note that, to further minimize bias, both ROBUST and UNCLEAR objects were retained 
in evaluating these average values of $\beta$, but in practice the $J_{140}$ 
cut ensures that virtually all objects are ROBUST, and rejection of the 
5 UNCLEAR objects at $z \simeq 7$, and the sole UNCLEAR object at $z \simeq 8$ 
does not significantly change these results. We note that the $M_{UV}$ 
bin centres quoted here refer to magnitudes based on 82\% of enclosed flux density
for a point source as detailed in Section 2.1, 
and are thus $\simeq 0.2$ mag fainter
than presumed total $M_{UV}$ for unresolved (or marginally-resolved) sources.
Finally, we note that the two-colour and power-law $\beta$ values for 
the individual objects are provided in Table A1.}

\begin{tabular}{llll}
\hline
$M_{UV}$ &  $\langle \beta \rangle$ ($J-H$) &  $\langle \beta \rangle$ (Power-law)&$\langle \beta \rangle$ (Power-law)\\
  & \phantom{0}Mean &  \phantom{0}Mean & \phantom{0}Weighted Mean\\     
\hline
&$z \simeq 7$\\
\hline
$-$19.5  &  $-$1.72$\pm$0.12 &  $-$1.81$\pm$0.12& $-$1.94$\pm$0.12\\
$-$18.5  &  $-$2.23$\pm$0.16 &  $-$2.08$\pm$0.15& $-$2.08$\pm$0.15\\
$-$17.5  &  $-$2.02$\pm$0.29 &  $-$2.08$\pm$0.26& $-$2.03$\pm$0.26\\
\\
\hline
&$z \simeq 8$\\
\hline
$-$19.5  &  $-$1.98$\pm$0.27 &  $-$2.03$\pm$0.17 &$-$1.93$\pm$0.17\\
$-$18.5  &  $-$1.96$\pm$0.27 &  $-$1.88$\pm$0.25 &$-$1.84$\pm$0.25\\
\\
\hline
&$z \simeq 9$\\
\hline
$-$18    &  $-$1.80$\pm$0.63 \\
\hline
\end{tabular}

 \end{center}
\end{table}

\subsection{Robust measurements at 
${\bmath {\bf z \simeq 7}}$ and ${\bmath {\bf z \simeq 8}}$}

In Fig. 1 we show the results of our new $J_{125}-H_{160}$ colour-based determinations of
$\beta$ for the galaxies in the new UDF12 samples at $z \simeq 7$ and $z \simeq 8$, plotted 
versus their UV absolute magnitudes
($M_{UV, AB} \equiv M_{1500}$).
To help provide dynamic range, the 
samples at $M_{UV,AB} < -19$ have been supplemented with $>8$-$\sigma$ 
objects from the UDF09P1 and UDF09P2 parallel WFC3/ACS fields; 
the individual measurements for all the UDF12, UDF09P1 and UDF09P2 sources 
utilised in this analysis are given in the Appendix in Table A1.

In this Fig. 1 we also plot the average values, along with standard errors
in the mean, for each 1-magnitude wide 
luminosity bin which contains $> 5 $ sources. These values are tabulated 
in Table 1. 
We emphasize that, in order to minimize colour bias in the selection 
process, the UDF12 galaxy samples used have been 
confined to those objects which are detected at 
$> 5\sigma$ in the $J_{140}$-band as described in Section 2.2. Moreover, 
to further minimize bias, both ROBUST and UNCLEAR objects were retained 
in evaluating these average values of $\beta$ (see Rogers et al. 2013), 
but in practice the $J_{140}$ 
cut ensures that virtually all objects are ROBUST, and rejection of the 
5 UNCLEAR objects at $z \simeq 7$, and the sole UNCLEAR object at $z \simeq 8$ 
does not significantly change these results. 

Our success in, for the first time, essentially eliminating any 
significant colour bias from these measurements at $z \simeq 7$ and $z \simeq 8$
is further confirmed below in Section 4, by the power-law $\beta$ determinations 
and associated end-to-end data simulations. Our new, robust results at $z \simeq 7$ 
confirm and extend the main conclusion of Dunlop et al. (2012), that there is, as yet, no evidence
for UV continua significantly bluer than $\beta \simeq -2$ in the currently 
detectable galaxy population at $z \simeq 7$. 

The results presented here at $z \simeq 8$ represent the first, meaningful 
and unbiased measurement of $\langle \beta \rangle$ for a significant sample of 
galaxies at this even earlier epoch, but we cannot probe to such faint absolute magnitudes
as at $z \simeq 7$. Nevertheless, over the available dynamic range  $-20 < M_{UV} < 18$
we clearly see no evidence for any significant change from $z \simeq 7$, with 
the average UV continuum slope $\langle \beta \rangle$ again consistent with $\beta = -2$
(the $z \simeq 8$ measurement remains more 
inaccurate simply due to smaller sample size).

\subsection{Preliminary measurements at ${\bmath {\bf z \simeq 9}}$}

In Fig. 2 we show the results of our attempt to determine $\beta$ 
for the 6 new galaxies we have uncovered in the HUDF at $8.5 < z < 10$,
as reported by Ellis et al. (2013). These measurements are necessarily based
on $J_{140} - H_{160}$ colour, and the individual values of 
$\beta$ thus derived are indicated by the red points in Fig. 2. The scatter
is very large, as expected given the photometric errors and the limitations
of equation (2) already discussed above. Nevertheless, since 
all of these objects have absolute 
UV magnitudes in the range $-18.5 < M_{UV} < -17.5$ we have proceeded 
to calculate the average $\langle \beta \rangle$ for a single bin 
centred at $M_{UV} \simeq -18$. This is shown by the black point,
which corresponds to $\langle \beta \rangle = -1.80 \pm 0.63$ (where the error 
is the standard error in the mean).

Such a measurement has not previously been possible at this redshift, 
and this first effort clearly yields a highly-uncertain 
result which should not be over-interpreted. 
Nevertheless, the current data provide no obvious
evidence that $\langle \beta \rangle$ at $M_{UV} = -18$ has changed 
dramatically between $z \simeq 7$ and $z \simeq 9$, with the average 
value still fully consistent with $\beta= -2$.

\section{Power-law measurements and data analysis simulations}

We now proceed to determine $\beta$ using the 
power-law fitting method as explored and optimized in Rogers et al. (2013). 
We have also performed a set of end-to-end data-analysis 
simulations, starting with the injection of sources into the real UDF12 
images, in order to quantify any remaining residual bias in our derived 
average values of $\langle \beta \rangle$. We first describe 
these simulations, before proceeding to summarize the results.

\subsection{Source injection, retrieval and measurement simulations}

Our simulations begin by defining a distribution of 
UV slopes. For the present study we adopt 
a delta function at $\beta=-2$ as our reference model, 
but also consider `top hat' distributions 
of various widths, as discussed below in Section 4.3.

Next, we create an input catalogue of galaxies 
with $\beta$ values drawn from the defined distribution, 
redshifts in the range $6<z<9$, and absolute magnitudes 
spanning $-22<M_{UV}<-16$ (with the relative number density of  
objects at different magnitudes governed by the latest $z=7$ luminosity function of 
McLure et al. 2013). A model SED is then created for each galaxy, incorporating the 
intrinsic colour, the IGM attenuation of flux blueward of the
Lyman break, the redshifting of the spectrum into the observed frame, 
and 
then cosmological dimming. Empirical PSFs are then created with 
broad-band flux-densities based on the model SEDs (in practice, 
the PSFs are set to zero in the $B_{435}, V_{606}, i_{775}$ bands 
where the flux is entirely attenuated). The PSFs are then inserted into the real
multi-wavelength UDF12 images, avoiding existing bright sources and regions of 
high rms noise 
where real candidates would be discarded. 

Objects are then reclaimed 
using SExtractor; we accept only objects lying within 2-pixels of an input 
PSF centre, and then perform aperture photometry 
on these objects in exactly the same way as for the real galaxies.
Photometric
redshifts are then obtained using Le Phare (Arnouts et al. 1999; Ilbert et al. 
2006) with the same BC03 models used in the real data analysis. 
We adopt an identical selection function to that used for the real data, and 
measure absolute magnitudes with the same synthetic filter on the
best-fitting BC03 model. Following Rogers et al. (2013), UV continuum slopes are 
obtained by performing a power-law fit to the $J_{125}, J_{140}, H_{160}$ 
flux densities 
($Y_{105}, J_{125}, H_{160}$ in the parallel fields,
accounting for partial attenuation of the $Y$-band by the Lyman break 
where appropriate).

\subsection{Results at ${\bmath {\bf z \simeq 7}}$}
In Fig. 3 we show the results of our power-law analysis for the galaxies 
at $z \simeq 7$, split into the same three luminosity bins as in Fig. 1, 
and this time, for completeness, showing results for both the FULL (upper row)
and ROBUST-source only (lower row) samples. 
The grid of power-law models fitted to the 
WFC3/IR photometry
extends over a deliberately very large range, $-8<\beta<5$, 
to ensure that the {\it observed} $\beta$ distribution is not 
artifically truncated. However, in practice (primarily due 
to the $J_{140}$ significance cut)
the sample studied here has photometry of sufficient quality 
that no object yields a measured $\beta<-4$.

The grey histograms in Fig. 3 
show the distribution of the power-law derived $\beta$ values 
in each bin, and the small squares with error bars 
indicate the average $\langle \beta \rangle$
values (and associated standard errors). The results derived from the 
real data are 
compared here with the results from our reference simulation in which every 
fake galaxy is assigned $\beta = -2$ before being inserted into the UDF12 
imaging; the red histograms indicate the distribution of power-law
$\beta$ values retrieved from the simulations, with the red points indicating 
the corresponding average and standard error in each luminosity bin. The red points 
thus offer a measure of the bias in our measurements of average $\langle \beta 
\rangle$ which can be seen to be negligible for both the FULL and 
ROBUST samples. As expected, it can be seen that confining the sample
to ROBUST sources results in the removal of a few of the redder
galaxies in the faintest magnitude bin, but because the number of UNCLEAR 
sources is so small, the results are essentially unchanged (especially
when measured relative to simulation expectation, which also moves slightly
blueward in the ROBUST-source simulations).

The final values given for the power-law determination of 
$\langle \beta \rangle$ at $z \simeq 7$ 
in Table 1 
are taken from the FULL-sample analysis shown in the upper row
of Fig. 3, 
and are calculated relative to the simulated values (to correct for any small 
residual bias). As long as the appropriate correction is applied, the results
are essentially identical if they are derived from the ROBUST-source 
only analysis presented in the lower row. Within the errors, all three 
luminosity bins at $z \simeq 7$ are clearly consistent with $\beta = -2$, 
with a best-estimate of $\langle \beta \rangle = -2.1$ in the fainter 
two bins. Reassuringly, the power-law estimates are fully
consistent with the $J_{125} - H_{160}$ colour-based measurements
presented in Section 3 (see Table 1 for details and errors, and Table A1 
for individual object measurements).

\begin{figure}
\epsfig{file=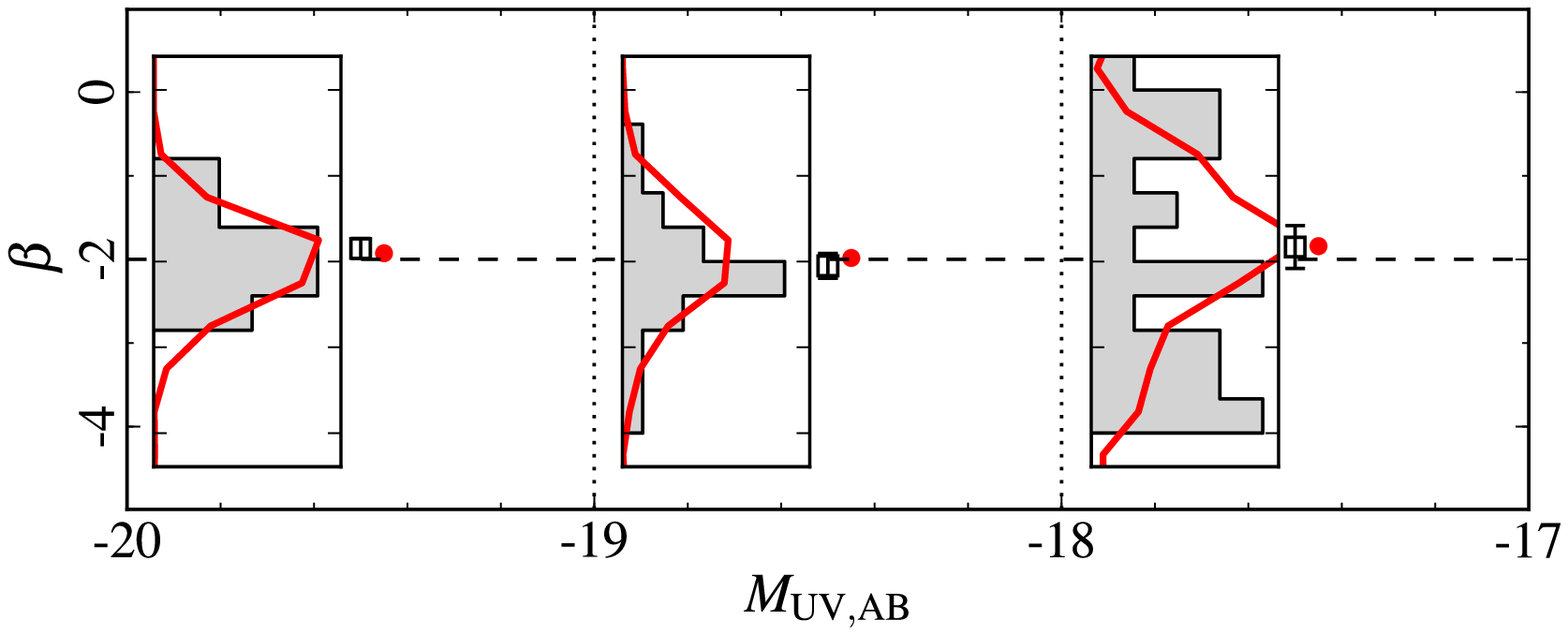, width=9.0cm,angle=0}
\epsfig{file=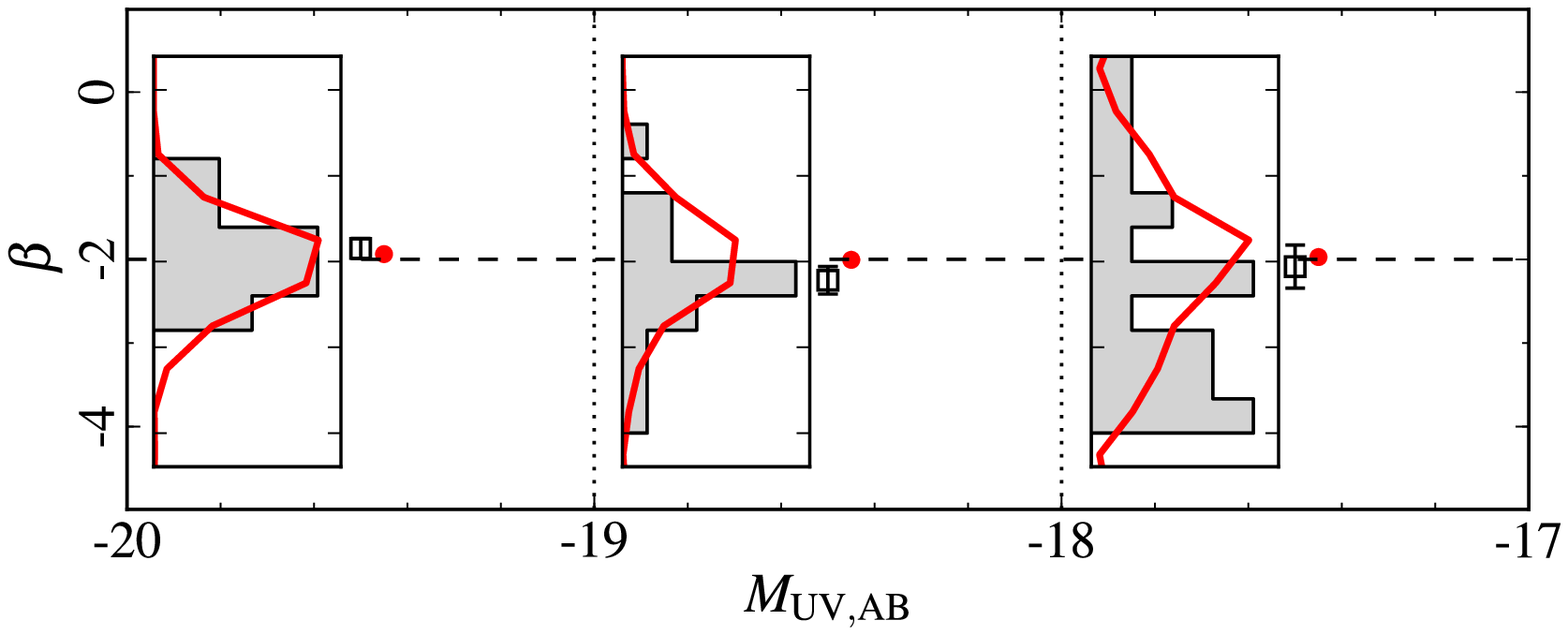, width=9.0cm,angle=0}

\caption{The distribution 
of individual power-law $\beta$ measurements at $z \simeq 7$, along 
with average values, $\langle \beta \rangle$ (and standard errors),
plotted against UV absolute magnitude.
Results are shown for all sources (upper row), and 
for ROBUST sources only (lower row). Simulations shown in red are based on 
2000 galaxies inserted with $\beta=-2$. The data from 
UDF12 are shown in grey/black.
The data in the brightest two bins have been supplemented with a few sources 
from the two UDF09 parallel fields as discussed in the text and in the caption to 
Fig. 1.}

\end{figure}

\begin{figure}
\epsfig{file=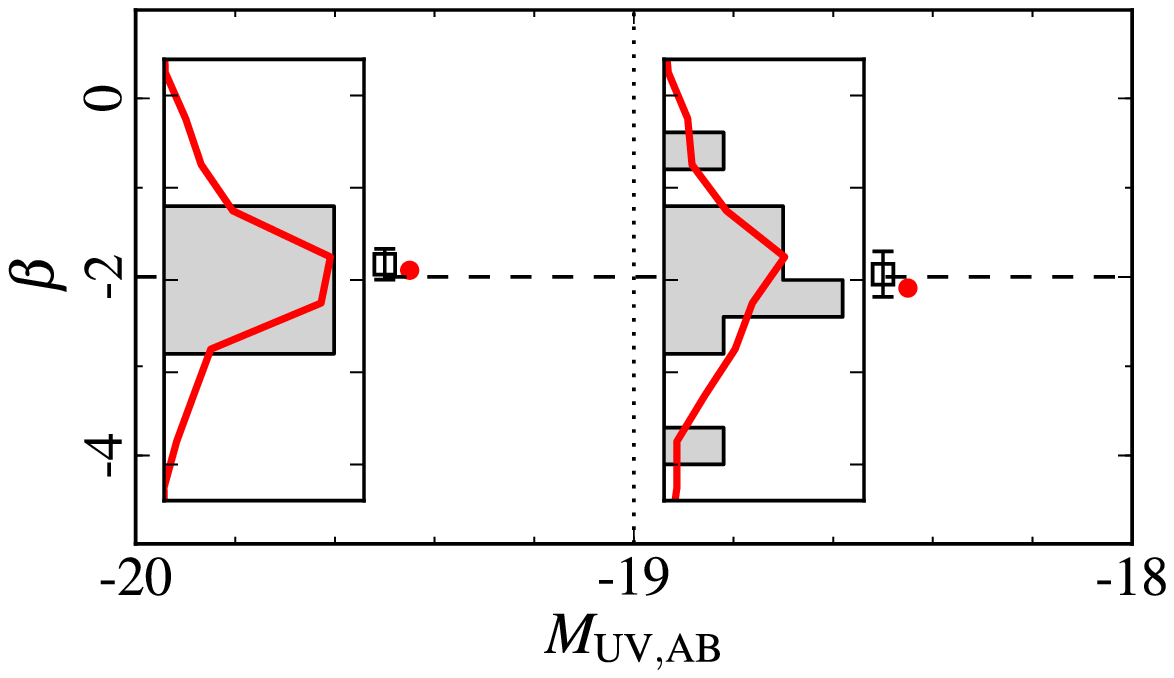, width=6.45cm,angle=0}
\epsfig{file=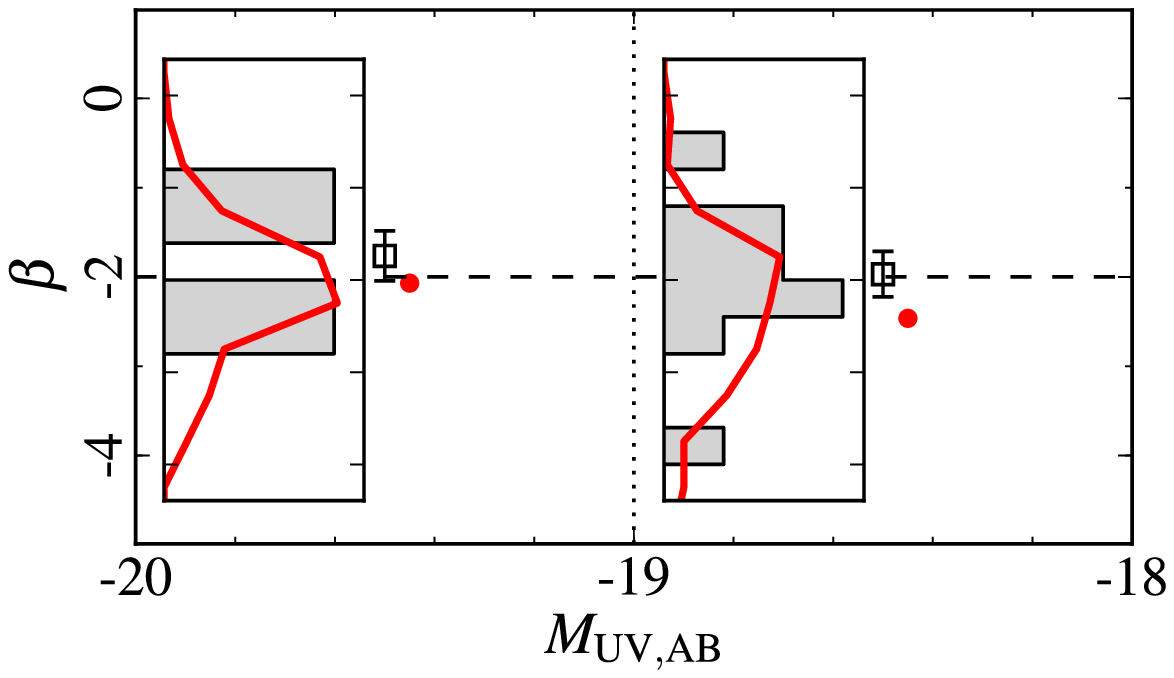, width=6.45cm,angle=0}

\caption{
The distribution 
of individual power-law $\beta$ measurements at $z \simeq 8$, along 
with average values, $\langle \beta \rangle$ (and standard errors)
plotted against UV absolute magnitude.
Results are shown for all sources (upper row), and 
for ROBUST sources only (lower row). 
Simulations shown in red are based on 
2000 galaxies inserted with $\beta=-2$. The data from 
UDF12 are shown in grey/black.
The faintest bin shown for the $z \simeq 7$ sources in Fig. 3 
only contains two sources in our $J_{140}$-thresholded $z \simeq 8$
sample, and so we do not attempt to 
show results at $M_{UV} \simeq -17.5$ here.
The data in the brighter bin have been supplemented with a few sources 
from the two UDF09 parallel fields,
as discussed in the text and in the caption to 
Fig. 1.}

\end{figure}

\subsection{Results at ${\bmath {\bf z \simeq 8}}$}

In Fig. 4 we show our power-law $\beta$ determinations at $z \simeq 8$. The 
values derived from the real data are again shown by the grey histograms, 
with the average and standard error indicated by the black squares with error bars.
Similarly, the corresponding
results for the $\beta = -2$ simulation are indicated in red. At $z \simeq 8$
the $J_{140}$ significance threshold leaves only two galaxies fainter than 
$M_{UV} = -18$ so, as in Fig. 1, we limit our analysis to the two brighter 
bins. The samples are smaller, and so the corresponding random errors
are larger, but again it can be seen that the values of $\langle \beta 
\rangle$ derived from the data are consistent with $\beta = -2$ in both
luminosity bins, and the blue bias implied from the simulations is relatively 
modest (although it is slightly larger if only ROBUST objects are 
retained, as expected). 

As at $z \simeq 7$, 
the final results for the power-law determination of $\langle \beta \rangle$
at $z \simeq 8$ given in Table 1 are derived from the FULL-sample analysis shown in the upper
row of Fig. 4, calculated relative to the simulated values. 
Again, within the errors, both
luminosity bins at $z \simeq 8$ are clearly consistent with $\beta = -2$,
and the power-law estimates are fully
consistent with the $J_{125} - H_{160}$ colour-based measurements
presented in Section 3 (see Table 1, and Table A1 
for individual object measurements).

\subsection{Trends with ${\bmath {\bf z}}$, ${\bmath {\bf M_{UV}}}$, and evidence for scatter}

Our derived average values of $\langle \beta \rangle$ reveal no evidence
for a significant trend with redshift over the limited redshift 
range explored here, $7 < z < 9$. However, it must be remembered 
that we are currently unable to explore the faintest absolute magnitude bin 
studied at $z \simeq 7$ at higher redshift.

Our results also do not yield any significant evidence for a 
relation between $\langle \beta \rangle$ and $M_{UV}$ at a given redshift,
although again the available dynamic range is limited, and a full exploration 
of this issue is deferred to a future paper including results from
brighter larger-area surveys.

The one suggestive result that does merit additional 
scrutiny here is the apparent excess scatter seen in $\beta$ at the faintest
absolute magnitudes probed at $z \simeq 7$. 
Specifically, in the lowest-luminosity bin  
plotted in Fig. 3 ($z\simeq7, M_{UV}>-18$) 
it appears that our simulation 
(red line in the histogram) does not replicate the 
observed scatter in $\beta$ (grey region) as successfully as 
at brighter magnitudes. It is of interest to attempt to 
quantify the statistical
significance of this effect, as a growth in the intrinsic 
scatter in $\beta$ with decreasing luminosity might be expected if, 
for example, the faintest galaxy samples
begin to include a significant number of very young, metal-poor 
objects.

To do this we have expanded our simulations (beyond 
a single value of $\beta = -2$) to explore a  
variety of intrinsic $\beta$ distributions. In particular, 
we considered alternative top-hat distributions for
the input values of $\beta$ in order to assess 
whether a wider intrinsic distribution 
can provide a significantly improved fit to the data in this faintest bin.
To determine the statistical significance of our results, we used 
a K-S test, as illustrated in the comparison of the simulated and observed
cumulative $\beta$ distributions presented in Fig. 5.
For clarity we restrict Fig. 5 to only 3 alternative input models, although
for consistency we again show results for the full  sample, and for ROBUST
sources only. From the K-S test significance values given in Fig. 5 
it can be seen that the $\beta = -2$ simulation in fact continues 
to provide a perfectly acceptable description of the data. Unsurprisingly,
a wider intrinsic distribution can provide an improved fit, although 
the highest significance values are achieved if this distribution
remains centred on a value close to $\beta = -2$ (consistent 
with our results for $\langle \beta \rangle$). Thus, while it is clear that 
we cannot rule out the possibility that our galaxy sample contains some objects
with UV slopes as blue as $\beta \simeq 3$ (see Table A1), the current data 
certainly do not require any significant intrinsic scatter 
(even in this well-populated luminosity bin). For now, therefore, 
reliable conclusions can only be drawn on the basis of population-averaged values, 
$\langle \beta \rangle$.
 
\begin{figure}
\centerline{\epsfig{file=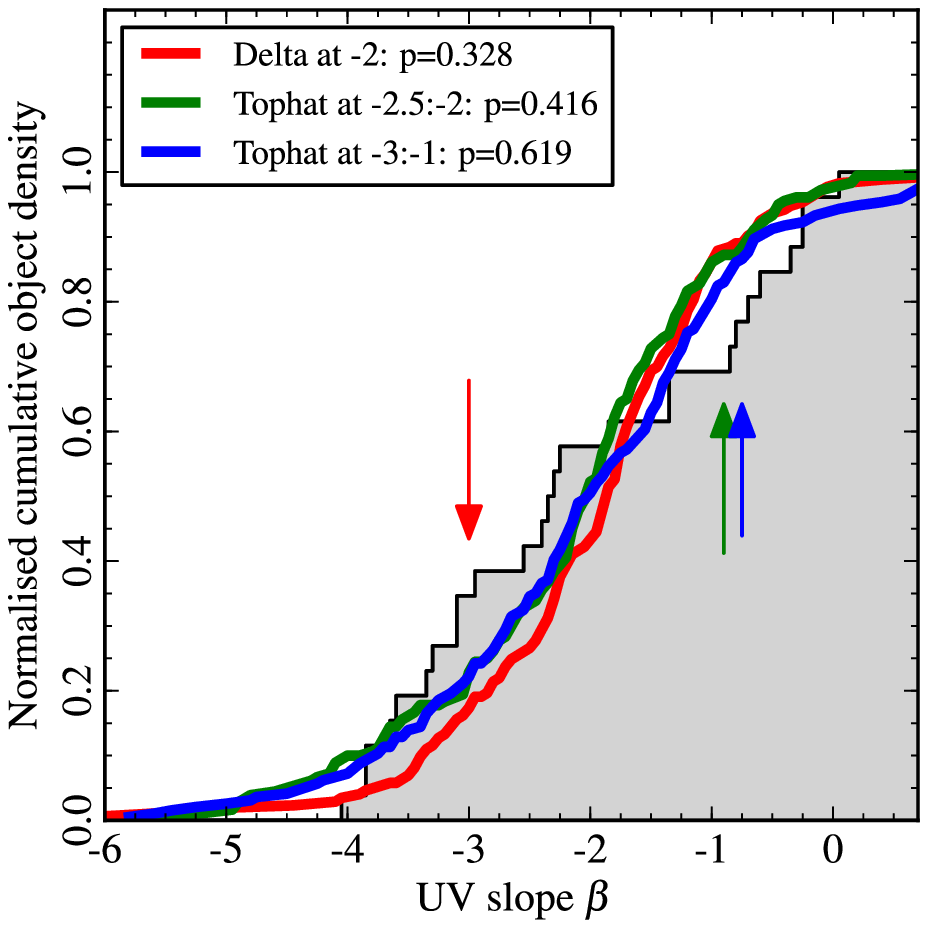,width=7.0cm,angle=0}}
\centerline{\epsfig{file=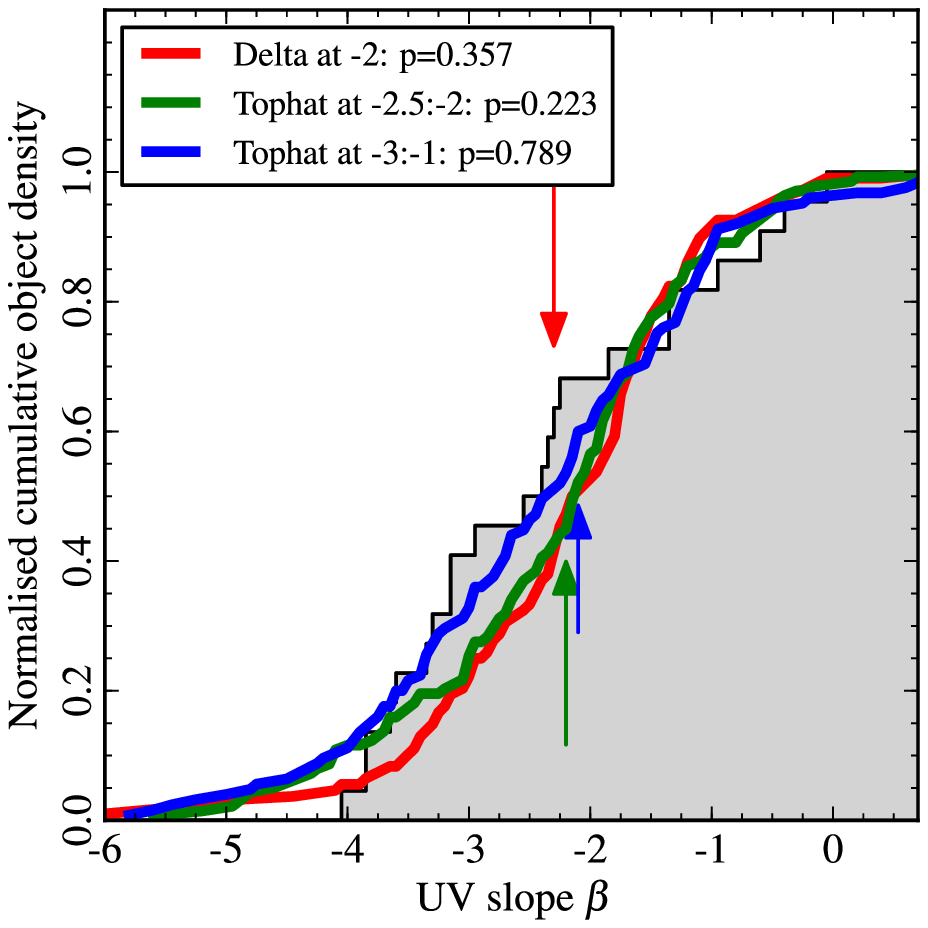,width=7.0cm,angle=0}}
\caption{A comparison of the distribution 
of $\beta$ values derived for the real galaxies in the 
faintest luminosity bin probed here at $z \simeq 7$ ($-18<M_{UV}<-17$),
with those predicted by alternative models based on different 
assumed intrinsic distributions of $\beta$. 
The upper panel shows all sources, while the lower panel
contains ROBUST sources only. The grey regions show the cumulative 
distributions of $\beta$ as derived from the data, 
while the coloured lines show the mock cumulative distributions
as produced by the output from each alternative simulation. The 
significance ($p$)
values for each model (under the null hypothesis that the real and simulated 
distributions are drawn from the same underlying distribution), 
are given in the top-left corner of each panel. 
Arrows show where the maximum deviation between the data and each 
simulation occurs, with the length of the arrow equal to the deviation.}
\end{figure}

\section{Discussion}

\subsection{Comparison with previous results}
At $z \simeq 7$ our results can be compared with the recent work of Bouwens et al. (2012)
and Finkelstein et al. (2012), and with our own previous study in Dunlop et al. (2012) 
which (as with the investigations by Bouwens et al. 2010b, Finkelstein et al. 2010, and Wilkins et al. 2011) was 
based purely on the portion of UDF09 WFC3/IR imaging obtained prior to the end of 2009.

Based on the complete UDF09 dataset, at $z \simeq 7$, Bouwens et al. (2012) reported a measurement for $\langle 
\beta \rangle$ at $M_{UV} \simeq -18.25$ of $\langle \beta \rangle = -2.68 \pm 0.19 \pm 0.28$, with the first
error representing the random error, and the second the estimated systematic uncertainty (albeit presumably, in practice,
not symmetric). This measurement 
is larger (redder) than the original Bouwens et al. (2010b) measurement of $-3.0 \pm 0.2$ in the same luminosity 
bin, with the change being due to the availability of the final UDF09 dataset, improved assessment of bias,
and (possibly) the removal of the $J_{125}$ flux-density threshold in the galaxy-selection process
(see Rogers et al. 2013). Clearly, however, the Bouwens et al. (2012) result still remains bluer 
than that reported here at comparable absolute magnitudes, albeit the two can be reconciled within the 
errors (especially if the estimated systematic error is applied to the red, moving 
$\langle \beta \rangle$ to $\simeq -2.40$). However, the Bouwens et al. (2012) results 
could easily still be biased blue in the faintest
bin, as they did not have the advantage of the deeper photometry from UDF12 exploited here, nor could the $J_{140}$ 
significance threshold be applied to assist in the selection of unbiased, secure sub-samples of galaxies.
Circumstantial evidence that the faintest measurement of $\langle \beta \rangle$ of Bouwens et al. (2012) 
remains biased to the blue is offered by the fact that the $\langle \beta \rangle$ value for their next brightest
bin is much redder, at $\langle \beta \rangle = -2.15 \pm 0.12 \pm 0.28$, in excellent agreement with our 
own results at comparable magnitudes.

Finkelstein et al. (2012) have also reported a move to redder values of $\langle \beta \rangle$ for the faintest 
galaxies at $z \simeq 7$ as found in the final UDF09 dataset compared to the original measurements made by 
Finkelstein et al. (2010). Specifically, Finkelstein et al. (2010) reported $\langle \beta \rangle = -3.07 \pm 0.51$, 
while Finkelstein et al. (2012) reported a median value of $\langle \beta \rangle = -2.68^{+0.39}_{-0.24}$ which becomes 
$\simeq -2.45$ after bias correction. Given the errors, clearly this result, 
while still somewhat bluer, can be reconciled with 
our own, now more accurate measurements.

In Dunlop et al. (2012) we aimed to highlight the dangers of the potential for blue bias in the 
early measurements of $\langle \beta \rangle$ made in the immediate aftermath of the first discovery of faint $z \simeq 7$ galaxies with WFC3/IR.
We utilised only the first epoch of the UDF09 dataset, and confined our attention to
$>8$-$\sigma$ sources, and therefore did not report a robust result for absolute magnitudes 
as faint as  $M_{UV} \simeq -18.5$ at $z \simeq 7$. We did, however, report $\langle \beta \rangle = -2.12 \pm 0.13$ 
at $M_{UV} \simeq -19.5$ for $z \simeq 7$ galaxies (in good agreement with the results from Bouwens et al. and 
Finkelstein et al. discussed above), and found 
$\langle \beta \rangle = -2.14 \pm 0.16$
at $M_{UV} \simeq -18.5$ for $z \simeq 5-6$. Clearly these results are consistent with the values in the corresponding 
luminosity bins presented here at $z \simeq 7$, confirming the apparent stability of $\langle \beta \rangle$ 
for the UV-selected population in this high-redshift regime (including the lack of any obvious redshift or luminosity dependence).

Finally, we note that Finkelstein et al. (2012) did attempt a measurement of $\langle \beta \rangle$ at $z \simeq 8$, and found 
$-2.00 \pm 0.32$, although this preliminary measurement was not deemed trustworthy enough for inclusion in the abstract (in part 
because it was substantially redder than at $z \simeq 7$). This result is in fact in very good agreement with the  
new, more robust determination at $z \simeq 8$ presented here, and indeed seems less surprising given our final results at $z \simeq 7$.

 \begin{figure}
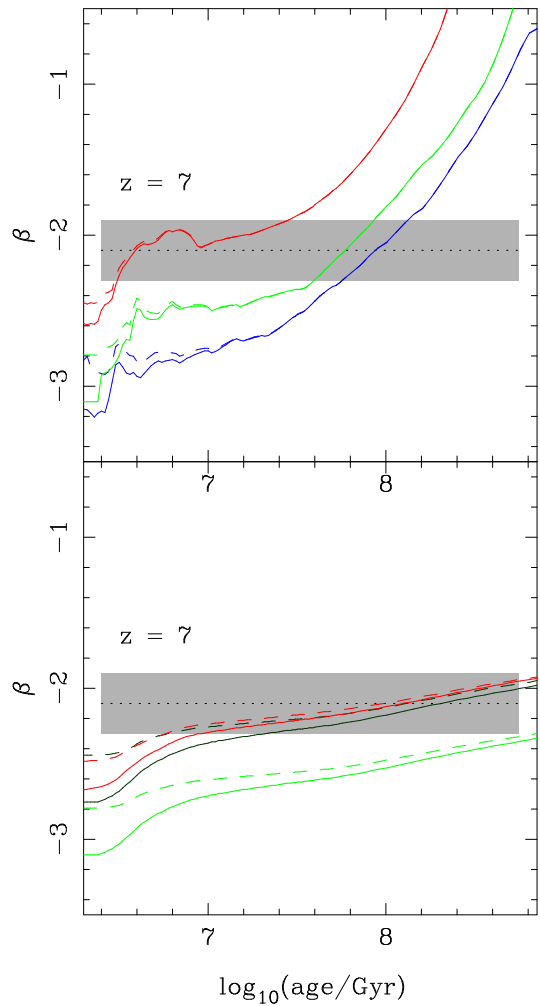

\centerline{\epsfig{file=fig6a.eps,width=7.0cm,angle=0}}
\vspace*{-0.47cm}
\centerline{\epsfig{file=fig6b.eps,width=7.0cm,angle=0}}
\caption{A comparison of our observed average value of $\langle \beta \rangle$ 
for galaxies 
with $M_{UV} \simeq -18$ at $z \simeq 7$ with the predicted age-dependence  
of $\beta$ from alternative stellar population models with and without nebular emission
(with age only plotted up to the age of the Universe at $z \simeq 7$).
In both panels our most robust data point is 
indicated by the horizontal dotted line, with the $1$-$\sigma$ uncertainty (standard error) 
shown by the surrounding grey shaded band. 
In the {\it upper panel} the solid lines are derived from 
instantaneous 
starburst models, with metallicities of $Z_{\odot}$ (red), 0.2$Z_{\odot}$ (green), 
and 0.02$Z_{\odot}$ (blue), assuming zero nebular emission (i.e. $f_{esc} = 1$). 
The dashed lines are produced by adding nebular continuum emission 
self-consistently (Robertson et al. 2010), 
assuming the extreme case $f_{esc} = 0$ (see text for details).
In the lower panel we show constant star-formation models, again with and without nebular emission,
but this time only for 
$Z_{\odot}$ (red) and 0.2$Z_{\odot}$ (green) models. The dark-green curves show the effect of adding 
modest
dust obscuration/reddening to the 0.2$Z_{\odot}$ (brighter green) model, ($\equiv A_V \simeq 0.1$ for SNII dust extinction,
or $\equiv A_V \simeq 0.2$ for the extinction law of Calzetti et al. 2000)
}
\end{figure}

\subsection{Physical interpretation}

Although we cannot exclude some intrinsic scatter in $\beta$, the analysis presented in 
Section 4.4 shows that we have, as yet, no evidence for it. What is clear is that our 
derived values of average $\langle \beta \rangle \simeq -2.1 \pm 0.2$ at $z \simeq 7$ clearly exclude 
the possibility that a {\it large subset} of galaxies in our 
sample have extreme values $\beta \simeq -3$, as anticipated 
from very young, very low-metallicity, dust-free stellar populations.

To illustrate this, and further explore the consequences of our measurements for the inferred 
physical properties of the currently-detectable galaxies, we show in Fig. 6 how our results 
compare with the values expected from stellar populations of different metallicity,  
nebular emission (related to ionizing escape fraction), dust reddening, 
and age (up to the age of the Universe
at $z \simeq 7$). In this figure our basic, most secure
result at $z \simeq 7$ and $M_{UV} \simeq -18$ 
(which is also consistent with our results at $z \simeq 8$ and $z \simeq 9$) 
is indicated by the horizontal dotted line, while the $1$-$\sigma$ uncertainty (standard error) is 
shown by the surrounding grey shaded band (which acceptable models should therefore intercept
at plausible ages).

The predictions of $\beta$ as a function of age shown in Fig. 6 have been 
produced using the BC03 evolutionary models. 
Nebular continuum emission has been added to the stellar-population
templates self-consistently, based on the
flux of Hydrogen-ionising photons predicted from each BC03 model
(using the code developed by Robertson et al. 2010). The nebular continuum includes
the emission of free-free and free-bound emission by H, neutral He and
singly-ionised He, as well as the two-photon continuum of H
(see the prescription given in Schaerer 2002).

In the upper panel of Fig. 6 we plot three alternative instantaneous 
starburst models, with metallicities equal to the solar value $Z_{\odot}$ (red), 0.2$Z_{\odot}$ (green), 
and 0.02$Z_{\odot}$ (blue). For each model we show the pure stellar prediction
(i.e. zero nebular emission $\equiv$ an ionizing photon escape 
fraction of unity, $f_{esc} = 1$) as a solid line, and the extreme alternative of maximum nebular 
contribution ($\equiv$ zero ionizing photon escape 
fraction, $f_{esc} = 0$) by the dashed line of the same colour. Not surprisingly for these burst models, 
the impact of the nebular continuum becomes negligible after $\simeq 10$\,Myr. During
the time period when it is significant, its impact is more pronounced the 
lower the adopted metallicity (see also figure 4 in Bouwens et al. 2010b).

From Fig. 6 we infer that our results are inconsistent with 
very young {\it and} very low metallicity models. Moreover, while $\beta \simeq -2.1$
can be produced by essentially any metallicity at a carefully-selected age, the 
speed with which the UV continuum reddens with age, coupled with the homogeneity 
of our results, argues strongly that the burst models illustrated in the upper panel 
(and in Bouwens et al. 2010b) are inappropriate, and in any case 
physically unrealistic.

A more natural assumption is that the galaxies selected by our rest-frame UV-selection technique
are forming stars quasi-continuously (at least on average, especially at these early times).
Therefore, in the lower panel of Fig. 6 we show predictions for constant star-formation
models. Again we show the $Z_{\odot}$ (red) and 0.2$Z_{\odot}$ (green) models, but this time (for clarity) 
omit the 0.02$Z_{\odot}$ model because, without dust-reddening, the 0.2$Z_{\odot}$ model is 
already too blue until an age of $\simeq 1$\,Gyr (and the inferred trend with further reduction
in metallicity is clear). Instead, we add a second 
version of the 0.2$Z_{\odot}$ model with modest
dust obscuration/reddening ($\equiv A_V \simeq 0.1$ for dust produced by Type-II 
supernovae (SNII),
or $\equiv A_V \simeq 0.2$ for the extinction law of Calzetti et al. 2000). This last curve 
(in dark green) illustrates the degeneracy between dust extinction and the assumed 
metallicity of the stellar population.

Clearly, these continuously star-forming models provide a much more plausible 
explanation of our results, and are capable of delivering the observed 
homogeneous value of $\beta$ without any requirement for fine tuning in age
(i.e. $\beta$ is little changed over the relevant timescale,
$\simeq 10$\,Myr to $\simeq 100$\,Myr). One is then left to choose 
between solar metallicity with very little room for any 
additional dust reddening, or moderately sub-solar metallicity stellar populations 
coupled with 
modest dust reddening. The degeneracies are clear, but the latter scenario 
is arguably more plausible, and in fact happens to correspond well with the 
physical properties predicted for the currently observable galaxies (i.e. $M_{UV} < -17$)
from
the cosmological galaxy-formation simulation discussed below.
Finally, we note that while, as expected, the contribution from nebular emission
persists to much longer times in these continually star-forming models, 
even the extremes adopted here of $f_{esc} = 1$ and $f_{esc} = 0$ yield
unobservably small differences in observed $\beta$. Thus, if quasi-continous
star-forming galaxies of modestly sub-solar metallicity are 
the correct interpretation of our results, there is currently no realistic 
prospect of estimating $f_{esc}$ from measurements of 
the UV continuum slope at $z \simeq 7$.
 
\subsection{Comparison with galaxy-formation model predictions}

It is instructive to 
compare our findings with the 
predictions of a state-of-the-art 
cosmological galaxy-formation simulation. 
The simulation used here has been recently described, 
and its basic observational predictions 
(e.g. luminosity function, mass function) 
verified by Dayal et al. (2013). Interested readers 
are referred to Maio et al. (2007, 2009, 2010) 
and Campisi et al. (2011) for complete details 
of the simulation, but the key details 
can be summarized as follows.

The simulation has been carried out using the 
TreePM-SPH code {\small {GADGET-2}} (Springel 2005), within the 
${\rm \Lambda}$CDM cosmology given in Section 1, and assuming a baryon density 
parameter $\Omega_{\rm b} = 0.04$, a 
primordial spectral index $n_s=1$ and a 
spectral normalisation $\sigma_8=0.9$.
The periodic simulation box has a comoving size of 
$10h^{-1}$\,Mpc and contains $320^3$ DM particles and, 
initially, an equal number of gas particles. The masses of the gas and 
DM particles are $3\times 10^5 h^{-1}\,{\rm M_{\odot}}$ and 
$2 \times 10^6 h^{-1}\,{\rm M_{\odot}}$, respectively. 

\begin{figure}
\centerline{\epsfig{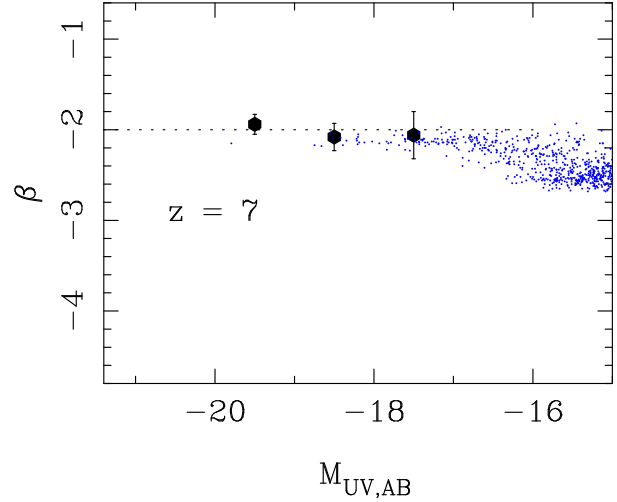}}
\caption{Comparison of our most accurate 
(bias-corrected power-law) measurements
of average $\langle \beta \rangle$ as a function of 
$M_{UV}$ at 
$z \simeq 7$ with the predicted $\beta$ values for 
individual galaxies as derived from the 
10\,Mpc cosmological galaxy-formation simulation described 
in Section 5.3 (see also Dayal et al. 2013).
The data plotted here are given in column 4 of Table 1.
The model predictions include the effects of dust, and therefore correspond to
the predicted {\it observed} values of $\beta$.}
\end{figure}

The code includes the molecular chemistry of 13 primordial species: 
${\rm e^-}$, ${\rm H}$, ${\rm H^-}$, ${\rm He}$, ${\rm He^+}$, 
${\rm He^{++}}$, ${\rm H_2}$, ${\rm H^+_2}$, ${\rm D}$, ${\rm D^+}$, 
${\rm HD}$, ${\rm HeH^+}$ (Yoshida et al. 2003; Maio et al. 2007, 2009), 
PopIII and PopII/I star formation according to the corresponding initial mass function 
(IMF; Tornatore, Ferrara \& Schneider 2007), gas cooling from resonant and fine-structure 
lines (Maio et al. 2007) and feedback effects (Springel \& Hernquist 2003). 
The runs track individual heavy elements (e.g. C, O, Si, Fe, Mg, S), and 
the transition from the metal-free PopIII to the metal-enriched 
PopII/I regime is determined by the underlying metallicity of the medium, 
$Z$, compared with the critical value of $Z_{crit} = 10^{-4}\,{\rm Z_{\odot}}$ 
(see Bromm \& Loeb 2003). If $Z<Z_{crit}$, a Salpeter IMF is 
used, with a mass range $100-500\,{\rm M_{\odot}}$; otherwise, a standard 
Salpeter IMF is used in the mass range $0.1-100\,{\rm M_{\odot}}$, and a 
SNII IMF is used for $8-40\,{\rm M_{\odot}}$ (Bromm et al. 2009; Maio et al. 2011).

The chemical model follows the detailed stellar evolution of each SPH particle. 
At every timestep, the abundances of different species are consistently derived using the 
lifetime function (Padovani \& Matteucci 1993) and metallicity-dependant stellar yields. 
The yields from SNII, AGB stars, SNIa and pair instability supernovae have been taken from 
Woosley \& Weaver (1995), van den Hoek \& Groenewegen (1997), Thielemann et al. (2003) 
and Heger \& Woosley (2002) respectively. Metal mixing is mimicked by smoothing metallicities 
over the SPH kernel and pollution is driven by wind feedback, which causes metal spreading 
over $\sim$\,kpc scales at each epoch (Maio et al. 2011). 

Galaxies are recognized as gravitationally-bound groups of at least 32 total 
(DM+gas+star) particles by running a friends-of-friends (FOF) algorithm, with a comoving 
linking-length of 0.2 in units of the mean particle separation. Substructures are identified 
by using the {\small SubFind} algorithm (Springel, Yoshida \& White 2001; Dolag et al. 2009) which 
discriminates between bound and non-bound particles. Of the galaxies identified in the 
simulation, we only use those that contain at least 10 star particles (at least 145 total 
particles) in our calculations at $z \simeq 7$. For each such well-resolved galaxy, we 
obtain the properties of all its star particles, including the redshift of, and 
mass/metallicity at formation.

Finally, the SED of each galaxy in the simulation snapshot at $z \simeq 7$ is 
calculated by assuming each star particle forms in a burst and then 
evolves passively. If, when a 
star particle forms, the metallicity of its parent gas particle is less than 
$Z_{crit} = 10^{-4}\,{\rm Z_{\odot}}$, we use the PopIII SED (Schaerer 2002); 
if a star particle forms out of a metal-enriched gas particle ($Z>Z_{crit}$), 
the SED is computed via the population synthesis code 
{\tt STARBURST99} (Leitherer et al. 1999), using its mass, 
stellar metallicity, and age. 
The composite spectrum for each galaxy is then calculated by summing the SEDs of all its 
star particles, and the intrinsic continuum luminosity, $L_c^{int}$, 
is calculated at $\lambda_{rest} = 1500$\,\AA.
We also self-consistently compute the dust 
mass and attenuation for each galaxy in the 
simulation box assuming type II SN (SNII) to be the main dust producers
(Maiolino et al. 2006; Stratta et al. 2007). The dust mass is converted 
into an optical depth to UV photons assuming the dust is made of 
carbonaceous grains spatially distributed as the gas 
(see Dayal et al. 2010 and Dayal \& Ferrera 2012
for complete details of this calculation). The predicted
observed UV luminosity is then $L_c^{obs} = L_c^{int} \times f_c$, 
where $f_c$ is the fraction of continuum photons that escape unattenuated 
by dust. The intrinsic value of $\beta$ ($\beta_{int}$) 
is calculated by fitting 
a power-law through the intrinsic SED of each galaxy over $\lambda_{rest} 
= 1500-3000$\AA\, with predicted `observed' $\beta$ values 
derived by repeating the 
power-law fitting after taking into 
account the dust enrichment and applying the SN extinction curve 
(Bianchi \& Schneider 2007).

In Fig. 7 we show  
the final predicted `observed' individual values of $\beta$ for galaxies 
in this  simulation at $z \simeq 7$, plotted 
against absolute UV magnitude, and compared against 
our best measurement of the actual (weighted) 
mean $\langle \beta \rangle$ values 
at $z \simeq 7$, as given in column 4 of Table 1.
Clearly the predictions of this simulation are in good 
agreement with our new results in a number of ways. First, 
within the magnitude range probed by our data, the average value 
of $\beta$ is basically exactly as predicted. The predicted 
intrinsic scatter in $\beta$ 
at these `brighter' magnitudes is also rather small, 
perfectly consistent with the analysis presented in Section 4.3. 
Moreover, the galaxies at $M_{UV} < -17$ in the simulation have 
moderately sub-solar metallicities (and hence intrinsically
blue $\beta$ values), but their chemical enrichment history is 
associated with enough dust to redden the observable values to 
$\beta$ values entirely consistent with our average results. 
Interestingly, this is essentially the same as one of the scenarios
we already considered in Section 5.3 as a possible explanation 
of our observed $\beta$ values. Finally, the simulation does 
not yield a significant $\beta - M_{UV}$ relation over the 
magnitude range probed in the current study, but does reinforce
the expectation that the distribution should broaden towards 
substantially bluer values of $\beta$ at very faint magnitudes
(as a significant population of low-metallicity, dust-free objects
finally emerges).

The agreement shown in Fig. 7 may be fortuitous,
and a full comparison with alternative galaxy-formation 
model predictions is beyond the scope of this paper. 
Nevertheless, Fig. 7 (and 
the associated physical properties of the simulated galaxies) 
does provide an interesting perspective on our findings.
In particular, it shows that an interpretation 
of our results in terms of only moderately 
sub-solar metallicities, coupled with modest dust reddening, is at 
least physically plausible within the magnitude range probed to date,
and that the long-anticipated 
population of really metal-poor dust-free objects 
possibly lies at much fainter magnitudes than originally 
suggested by the early results of Bouwens et al. (2010b).
In addition, it serves to remind us that the theoretically-predicted scatter 
in $\beta$ at these epochs is on a completely different scale 
to that seen in our raw individual galaxy $\beta$ distributions, 
the latter being still utterly  
dominated by the impact of photometric errors (compare, for example,
Fig. 7 and Fig. 1). Nevertheless, Fig. 7 also provides 
strong continued motivation for pushing the robust 
measurement of $\langle \beta \rangle$ 
to ever fainter galaxy luminosities in the young Universe.
 
\subsection{Implications for reionization}
A full analysis of the implications of the combined
results of the UDF12 programme for our understanding of cosmic
reionization is presented by Robertson et al. (2013).
This analysis will necessarily draw on the new luminosity-function
measurements presented by McLure et al. (2013) and
Schenker et al. (2013), and on the new measurements
of $\langle \beta \rangle$ presented here, because determining the
ability of the emerging galaxy population at $z = 6 - 9$ to reionize the
Universe requires knowledge not only of the number density of the galaxies,
but also information on their ability to supply the required ionizing photons.
As discussed in Section 1, the ability of a given galaxy to contribute to the
ionization of the surrounding inter-galactic medium
depends on its luminosity, the age and metallicity of its stellar population,
and the escape fraction of ionizing photons

Unfortunately, the relatively modest values of $\langle \beta \rangle$
found in the present study do not easily lend themselves to straightforward
interpretation. In particular, Fig. 6 suggests that there is little prospect
of using our new measurements of $\langle \beta \rangle$ to set new
constraints on the ionizing-photon escape fraction, $f_{esc}$, without
significant additional information (e.g. a meaningful estimate of the
contribution of nebular emission lines to the rest-frame optical colours,
as measured by {\it Spitzer} IRAC; e.g., Stark et al. 2013; Labb\'{e} et al. 2013).

On the other hand, the degeneracy between metallicity and dust obscuration
is somewhat less of an issue for reionization calculations than 
for calculations of cosmic star-formation rate density, as the former
concerned only with the UV photons which 
survive to exit a galaxy and potentially
contribute to cosmic reionization.
Moreover, our finding that $\langle \beta \rangle$ remains
close to $\beta = -2$ at $z \simeq 7, 8$ and even $z \simeq 9$ suggests
that the galaxies detected to date already contain relatively mature,
metal-enriched stellar populations, lending support to a picture in which
star-formation (and hence cosmic reionization) commenced at significantly
higher redshifts.

\section{Conclusions}

We have used the new ultra-deep, near-infrared imaging of the HUDF 
provided by our UDF12 HST WFC3/IR imaging campaign 
to explore the rest-frame UV properties of galaxies at redshifts 
$z > 6.5$. In this study we have exploited 
the final multi-band WFC3/IR 
imaging (UDF12+UDF09) to select deeper and more reliable 
galaxy samples at $z \simeq 7$, $z \simeq 8$, and $z \simeq 9$, and 
to provide improved photometric redshifts. Most importantly,
we have used the enhanced dataset to provide more accurate photometry,
and to base galaxy selection primarily on the new $J_{140}$ imaging, which 
exerts minimal influence on the derivation of the UV spectral index $\beta$ 
($f_{\lambda}~\propto~\lambda^{\beta}$) from 
either $J_{125}-H_{160}$ colour (e.g. Bouwens et al. 2010b; Dunlop et al. 2012), 
or $J_{125}+J_{140}+H_{160}$ power-law fitting
(as advocated by Rogers et al. 2013). Our main results are as follows.

\vspace*{0.1in}

\noindent
{\bf i)} We have produced the first robust and unbiased measurement of the
average UV power-law index, $\langle \beta \rangle$, 
($f_{\lambda}~\propto~\lambda^{\beta}$) for faint galaxies at $z \simeq 7$, 
finding $\langle \beta \rangle = -2.1 \pm 0.2$ at $z \simeq 7$ for galaxies 
with $M_{UV} \simeq -18$. This result means that the faintest galaxies 
uncovered to date at this epoch have, {\it on average}, UV colours 
no more extreme than those displayed by the bluest star-forming galaxies 
found in the low-redshift Universe.

\vspace*{0.1in}

\noindent
{\bf ii)} We have made the first meaningful measurements of 
$\langle \beta \rangle$ at $z~\simeq~8$, 
finding a similar value, $\langle \beta \rangle = -1.9 \pm 0.3$.

\vspace*{0.1in}

\noindent
{\bf iii)} We have offered a tentative first estimate of 
$\langle \beta \rangle$ at $z \simeq 9$ (based on the $J_{140}-H_{160}$ colours 
of the six galaxies in the redshift range $8.5 < z < 10$ reported in Ellis et al. 2013), 
and find $\langle \beta \rangle = -1.8 \pm 0.6$, essentially unchanged from 
$z \simeq 6-7$ (albeit highly uncertain).

\vspace*{0.1in}

\noindent
{\bf iv)} Finally, we have used careful end-to-end source injection+retrieval+analysis 
simulations to quantify any small residual biases in our measurements, and 
to test for any evidence for significant intrinsic scatter in the $\beta$ values 
displayed by the galaxies in the faintest luminosity bin which we can study at $z \simeq 7$.
While models including a range of $\beta$ values provide a modestly-improved 
description of the data,  we find that there is, as yet, no evidence for a significant {\it intrinsic} 
scatter in $\beta$ within our new $z \simeq 7$ galaxy sample. 

\vspace*{0.1in}

\noindent
Our results exclude the possibility that even our faintest galaxy samples 
contain a substantial population
of very low-metallicity, dust-free objects with $\beta \simeq -3$. Rather, 
our findings 
are most easily explained by a population of steadily star-forming
galaxies with either $\simeq$ solar metallicity and zero dust, or moderately
sub-solar ($\simeq 10-20$\%) metallicity with modest dust obscuration ($A_V
\simeq 0.1-0.2$). This latter interpretation
is consistent with the predictions of a state-of-the-art galaxy-formation 
simulation, which also suggests that a significant population
of very-low metallicity, dust-free galaxies with $\beta \simeq -2.5$ 
may not emerge until $M_{UV} > -16$, a regime 
likely to remain inaccessible until the JWST.

\section*{ACKNOWLEDGEMENTS}
JSD and PD thank their collaborators U. Maio and B. Ciardi for making available the SPH simulations utilised in Section 5.3.
JSD, PD, VW, RAAB, and TAT acknowledge the support of the European Research Council via the award of an Advanced Grant.
JSD and RJM acknowledge the support of the Royal Society via a Wolfson Research Merit Award, and a University Research Fellowship respectively. 
ABR and EFCL acknowledge the support of the UK Science \& Technology Facilities Council.
US authors acknowledge financial support from the Space Telescope Science Institute under 
award HST-GO-12498.01-A. 
SRF is partially supported by the David and Lucile Packard Foundation.
SC acknowledges the support of the European Commission through the Marie Curie Initial Training Network ELIXIR. 
This work is based in part on observations made with the NASA/ESA {\it Hubble Space Telescope}, which is operated by the Association 
of Universities for Research in Astronomy, Inc, under NASA contract NAS5-26555.
This work is also based in part on observations made with the {\it Spitzer Space Telescope}, which is operated by the Jet Propulsion Laboratory, 
California Institute of Technology under NASA contract 1407. 

{}

\appendix

\section{UV slope (${\bf \beta}$) measurements for individual objects}

In this appendix we provide a table of the key measured/derived properties (positions, photometric 
redshifts, absolute magnitudes, and UV slopes $\beta$) of the objects analysed 
to produce the plots and statistical averages in this paper. The sample summarised in Table A1
comprises the subset of the HUDF12 $z_{phot} = 6.4 - 8.4$ sample which survived the $J_{140}$ selection threshold described in 
Section 2.2, supplemented by a small number of brighter objects from the two UDF Parallel fields.

\begin{table*}
\caption{Positions, photometric redshifts, absolute UV magnitudes, and $\beta$ values for the objects 
analysed in this paper. We give both the simple $J-H$ colour determinations of $\beta$, and our 
preferred optimal power-law determinations, the latter with errors. We caution that, given the typical
sizes of these errors, individual measurements of $\beta$ must be regarded as highly uncertain. Sources
are ordered by photometric redshift (see McLure et al. 2013), with the source name in column 1 indicating 
whether the source lies within the UDF itself, or the parallel fields UDFPar1 and UDFPar2.}
\begin{center}
\begin{tabular}{l l l r r r r}
\hline
Name & RA(J2000) & Dec(J2000) & $z_{phot}$ & $M_{1500}$ & $\beta_{J-H}$ & $\beta_{\rm{PL}}\pm1\sigma$ \\
\hline
UDF12-3732-6420 &   03:32:37.32 &   $-$27:46:42.0   &   6.4 &   $-$17.4 &   $-$4.0  &   $-$3.6$\pm$1.5  \\ 
UDF12-3983-6189 &   03:32:39.83 &   $-$27:46:18.9   &   6.4 &   $-$17.7 &   $-$1.8  &   $-$1.8$\pm$1.1  \\ 
UDF12-3677-7536 &   03:32:36.77 &   $-$27:47:53.6   &   6.4 &   $-$18.8 &   $-$2.2  &   $-$2.1$\pm$0.7  \\ 
UDF12-3696-5536 &   03:32:36.96 &   $-$27:45:53.6   &   6.5 &   $-$17.3 &   $-$2.3  &   $-$2.3$\pm$1.2  \\ 
UDF12-4058-5570 &   03:32:40.58 &   $-$27:45:57.0   &   6.5 &   $-$17.8 &   $-$1.3  &   $-$1.3$\pm$1.1  \\ 
UDF12-3900-6482 &   03:32:39.00 &   $-$27:46:48.2   &   6.5 &   $-$17.9 &   $-$0.2  &   $-$0.3$\pm$0.7  \\ 
UDF12-3638-7163 &   03:32:36.38 &   $-$27:47:16.3   &   6.5 &   $-$18.5 &   $-$2.3  &   $-$2.3$\pm$0.6  \\ 
UDF12-4056-6436 &   03:32:40.56 &   $-$27:46:43.6   &   6.5 &   $-$18.5 &   $-$1.5  &   $-$1.9$\pm$0.6  \\ 
UDFPar2-0194-2033   &   03:33:01.94 &   $-$27:52:03.3   &   6.5 &   $-$19.3 &   $-$1.6  &   $-$1.7$\pm$0.3  \\ 
UDF12-3865-6041 &   03:32:38.65 &   $-$27:46:04.1   &   6.5 &   $-$17.6 &   $-$3.8  &   $-$3.6$\pm$1.2  \\ 
UDF12-4472-6362 &   03:32:44.72 &   $-$27:46:36.2   &   6.5 &   $-$17.9 &   0.8 &   0.4$\pm$1.1 \\ 
UDF12-4268-7073 &   03:32:42.68 &   $-$27:47:07.3   &   6.5 &   $-$18.1 &   $-$2.0  &   $-$2.0$\pm$0.6  \\ 
UDF12-3736-6245 &   03:32:37.36 &   $-$27:46:24.5   &   6.6 &   $-$17.6 &   $-$2.5  &   $-$2.5$\pm$1.1  \\ 
UDF12-4182-6112 &   03:32:41.82 &   $-$27:46:11.2   &   6.6 &   $-$17.8 &   $-$3.6  &   $-$3.5$\pm$0.7  \\ 
UDF12-4219-6278 &   03:32:42.19 &   $-$27:46:27.8   &   6.6 &   $-$18.9 &   $-$2.1  &   $-$2.1$\pm$0.6  \\ 
UDF12-4256-7314 &   03:32:42.56 &   $-$27:47:31.4   &   6.6 &   $-$19.3 &   $-$2.1  &   $-$2.2$\pm$0.6  \\ 
UDFPar2-0977-0485   &   03:33:09.77 &   $-$27:50:48.5   &   6.6 &   $-$19.1 &   $-$0.5  &   $-$0.8$\pm$0.3  \\ 
UDF12-3734-7192 &   03:32:37.34 &   $-$27:47:19.2   &   6.7 &   $-$17.8 &   $-$3.0  &   $-$3.0$\pm$1.1  \\ 
UDF12-3675-6447 &   03:32:36.75 &   $-$27:46:44.7   &   6.7 &   $-$17.9 &   $-$2.4  &   $-$2.4$\pm$0.7  \\ 
UDF12-4160-7045 &   03:32:41.60 &   $-$27:47:04.5   &   6.7 &   $-$18.3 &   $-$2.3  &   $-$2.3$\pm$0.6  \\ 
UDF12-3836-6119 &   03:32:38.36 &   $-$27:46:11.9   &   6.7 &   $-$18.5 &   $-$2.1  &   $-$1.6$\pm$0.7  \\ 
UDFPar2-1044-1081   &   03:33:10.44 &   $-$27:51:08.1   &   6.7 &   $-$18.5 &   $-$2.4  &   $-$2.1$\pm$0.6  \\ 
UDF12-3471-7236 &   03:32:34.71 &   $-$27:47:23.6   &   6.7 &   $-$17.4 &   $-$0.1  &   $-$0.2$\pm$0.9  \\ 
UDF12-4068-6498 &   03:32:40.68 &   $-$27:46:49.8   &   6.7 &   $-$17.5 &   $-$3.0  &   $-$2.9$\pm$1.1  \\ 
UDF12-3975-7451 &   03:32:39.75 &   $-$27:47:45.1   &   6.7 &   $-$17.7 &   $-$1.1  &   $-$1.3$\pm$1.1  \\ 
UDF12-3744-6513 &   03:32:37.44 &   $-$27:46:51.3   &   6.7 &   $-$18.8 &   $-$2.7  &   $-$2.7$\pm$0.7  \\ 
UDFPar1-5693-0509   &   03:32:56.93 &   $-$27:40:50.9   &   6.7 &   $-$18.7 &   $-$3.5  &   $-$2.6$\pm$0.5  \\ 
UDF12-3729-6175 &   03:32:37.29 &   $-$27:46:17.5   &   6.8 &   $-$17.9 &   $-$0.8  &   $-$0.7$\pm$0.8  \\ 
UDF12-3958-6565 &   03:32:39.58 &   $-$27:46:56.5   &   6.8 &   $-$18.7 &   $-$1.8  &   $-$1.8$\pm$0.6  \\ 
UDFPar2-0419-0314   &   03:33:04.19 &   $-$27:50:31.4   &   6.8 &   $-$20.1 &   $-$1.5  &   $-$1.5$\pm$0.3  \\ 
UDF12-3989-6189 &   03:32:39.89 &   $-$27:46:18.9   &   6.8 &   $-$17.8 &   $-$3.1  &   $-$3.0$\pm$1.0  \\ 
UDF12-4431-6452 &   03:32:44.31 &   $-$27:46:45.2   &   6.8 &   $-$18.5 &   $-$3.3  &   $-$3.3$\pm$0.6  \\ 
UDFPar2-0914-1531   &   03:33:09.14 &   $-$27:51:53.1   &   6.8 &   $-$18.5 &   $-$2.4  &   $-$2.4$\pm$0.4  \\ 
UDF12-3456-6494 &   03:32:34.56 &   $-$27:46:49.4   &   6.8 &   $-$17.6 &   $-$0.5  &   $-$0.5$\pm$1.0  \\ 
UDF12-4105-7156 &   03:32:41.05 &   $-$27:47:15.6   &   6.8 &   $-$18.7 &   $-$1.4  &   $-$0.9$\pm$0.6  \\ 
UDFPar1-0210-1463   &   03:33:02.10 &   $-$27:41:46.3   &   6.8 &   $-$19.3 &   $-$1.8  &   $-$1.9$\pm$0.3  \\ 
UDFPar2-0119-1134   &   03:33:01.19 &   $-$27:51:13.4   &   6.8 &   $-$19.1 &   $-$1.5  &   $-$1.5$\pm$0.3  \\ 
UDFPar1-5881-0386   &   03:32:58.81 &   $-$27:40:38.6   &   6.9 &   $-$18.5 &   $-$2.7  &   $-$2.7$\pm$0.5  \\ 
UDFPar1-0243-1313   &   03:33:02.43 &   $-$27:41:31.3   &   6.9 &   $-$19.3 &   $-$1.7  &   $-$2.0$\pm$0.3  \\ 
UDFPar2-0631-1217   &   03:33:06.31 &   $-$27:51:21.7   &   6.9 &   $-$18.7 &   $-$2.1  &   $-$2.1$\pm$0.5  \\ 
UDF12-4256-6566 &   03:32:42.56 &   $-$27:46:56.6   &   7.0 &   $-$20.0 &   $-$1.4  &   $-$1.4$\pm$0.6  \\ 
UDFPar2-0964-0508   &   03:33:09.64 &   $-$27:50:50.8   &   7.0 &   $-$20.3 &   $-$1.8  &   $-$2.3$\pm$0.3  \\ 
UDFPar2-0914-1555   &   03:33:09.14 &   $-$27:51:55.5   &   7.0 &   $-$19.5 &   $-$1.6  &   $-$2.6$\pm$0.3  \\ 
UDFPar2-0704-0555   &   03:33:07:04 &   $-$27:50:55.5   &   7.0 &   $-$19.3 &   $-$1.7 &   $-$1.7$\pm$0.3 \\
UDF12-4245-6534 &   03:32:42.45 &   $-$27:46:53.4   &   7.0 &   $-$17.1 &   0.4 &   $-$0.2$\pm$1.6  \\ 
UDF12-4071-7347 &   03:32:40.71 &   $-$27:47:34.7   &   7.0 &   $-$17.5 &   $-$2.4  &   $-$2.3$\pm$1.5  \\ 
UDF12-3431-7115 &   03:32:34.31 &   $-$27:47:11.5   &   7.0 &   $-$18.2 &   $-$0.6  &   $-$0.6$\pm$0.7  \\ 
UDF12-3825-6566 &   03:32:38.25 &   $-$27:46:56.6   &   7.0 &   $-$17.2 &   $-$0.3  &   $-$0.8$\pm$1.6  \\ 
UDF12-3853-7519 &   03:32:38.53 &   $-$27:47:51.9   &   7.0 &   $-$17.8 &   $-$3.4  &   $-$3.3$\pm$1.2  \\ 
UDFPar1-5897-0504   &   03:32:58.97 &   $-$27:40:50.4   &   7.0 &   $-$19.3 &   $-$1.7  &   $-$1.7$\pm$0.3  \\ 
UDFPar1-5959-1209   &   03:32:59.59 &   $-$27:41:20.9   &   7.0 &   $-$19.4 &   $-$2.0  &   $-$2.1$\pm$0.3  \\ 
UDFPar1-5670-1082   &   03:32:56.70 &   $-$27:41:08.2   &   7.0 &   $-$19.6 &   $-$1.9  &   $-$1.6$\pm$0.2  \\ 
UDFPar2-0540-1189   &   03:33:05.40 &   $-$27:51:18.9   &   7.0 &   $-$19.1 &   $-$2.3  &   $-$2.5$\pm$0.4  \\ 
UDFPar1-5969-0353   &   03:32:59.69 &   $-$27:40:35.3   &   7.1 &   $-$19.7 &   $-$2.3  &   $-$2.3$\pm$0.3  \\	
UDFPar1-5850-0239   &   03:32:58.50 &   $-$27:40:23.9   &   7.1 &   $-$19.2 &   $-$2.4  &   $-$2.5$\pm$0.4  \\ 
UDF12-4242-6243 &   03:32:42.42 &   $-$27:46:24.3   &   7.2 &   $-$18.1 &   $-$1.5  &   $-$1.6$\pm$0.6  \\ 
UDF12-3402-6504 &   03:32:34.02 &   $-$27:46:50.4   &   7.2 &   $-$18.2 &   $-$3.3  &   $-$3.1$\pm$0.7  \\ 
UDF12-4242-6137 &   03:32:42.42 &   $-$27:46:13.7   &   7.2 &   $-$17.8 &   $-$0.6  &   $-$0.8$\pm$1.2  \\ 
UDF12-4384-6311 &   03:32:43.84 &   $-$27:46:31.1   &   7.3 &   $-$17.7 &   $-$3.5  &   $-$3.3$\pm$1.3  \\ 
UDF12-3668-8067 &   03:32:36.68 &   $-$27:48:06.7   &   7.3 &   $-$17.9 &   $-$4.0  &   $-$3.6$\pm$1.2  \\ 
UDF12-3973-6214 &   03:32:39.73 &   $-$27:46:21.4   &   7.3 &   $-$18.2 &   $-$3.6  &   $-$3.6$\pm$0.8  \\ 
\end{tabular}
\end{center}
\label{default}
\addtocounter{table}{-1}
\end{table*}

\begin{table*}
\caption{cont.}
\begin{center}
\begin{tabular}{l l l r r r r}
\hline
Name & RA(J2000) & Dec(J2000) & $z_{phot}$ & $M_{1500}$ & $\beta_{J-H}$ & $\beta_{\rm{PL}}\pm1\sigma$ \\
\hline
UDFPar1-5575-1070   &   03:32:55.75 &   $-$27:41:07.0   &   7.3 &   $-$19.1 &   $-$0.9  &   $-$0.9$\pm$0.5  \\ 
UDF12-3313-6545 &   03:32:33.13 &   $-$27:46:54.5   &   7.3 &   $-$18.4 &   $-$2.2  &   $-$2.2$\pm$0.6  \\ 
UDFPar1-5973-1193   &   03:32:59.73 &   $-$27:41:19.3   &   7.4 &   $-$19.3 &   $-$1.7  &   $-$2.0$\pm$0.4  \\ 
UDF12-3885-7540 &   03:32:38.85 &   $-$27:47:54.0   &   7.5 &   $-$17.9 &   $-$2.2  &   $-$2.2$\pm$1.0  \\ 
UDF12-3931-6181 &   03:32:39.31 &   $-$27:46:18.1   &   7.5 &   $-$18.0 &   $-$4.0  &   $-$3.9$\pm$1.0  \\ 
UDF12-3880-7072 &   03:32:38.80 &   $-$27:47:07.2   &   7.5 &   $-$19.9 &   $-$1.4  &   $-$1.4$\pm$0.6  \\ 
UDFPar1-5939-2017   &   03:32:59.39 &   $-$27:42:01.7   &   7.5 &   $-$18.9 &   $-$1.0  &   $-$1.3$\pm$0.7  \\ 
\hline
UDFPar2-0376-1197   &   03:33:03.76 &   $-$27:51:19.7   &   7.5 &   $-$20.0 &   $-$1.5  &   $-$1.6$\pm$0.4  \\ 
UDFPar2-0378-1204   &   03:33:03.78 &   $-$27:51:20.4   &   7.5 &   $-$20.4 &   $-$1.1  &   $-$1.4$\pm$0.3  \\ 
UDF12-4470-6443 &   03:32:44.70 &   $-$27:46:44.3   &   7.7 &   $-$19.4 &   $-$1.8  &   $-$1.9$\pm$0.7  \\ 
UDF12-4288-6345 &   03:32:42.88 &   $-$27:46:34.5   &   7.7 &   $-$18.7 &   $-$1.4  &   $-$1.4$\pm$0.8  \\ 
UDF12-3722-8061 &   03:32:37.22 &   $-$27:48:06.1   &   7.7 &   $-$18.9 &   $-$1.9  &   $-$1.9$\pm$0.7  \\ 
UDF12-3952-7174 &   03:32:39.52 &   $-$27:47:17.4   &   7.7 &   $-$19.0 &   $-$0.7  &   $-$0.7$\pm$0.7  \\ 
UDFPar2-0464-0530   &   03:33:04.64 &   $-$27:50:53.0   &   7.7 &   $-$19.6 &   $-$2.1  &   $-$2.0$\pm$0.4  \\ 
UDF12-4309-6260 &   03:32:43.09 &   $-$27:46:26.0   &   7.7 &   $-$18.1 &   $-$2.2  &   $-$2.2$\pm$0.8  \\ 
UDF12-3939-7040 &   03:32:39.39 &   $-$27:47:04.0   &   7.7 &   $-$18.2 &   $-$1.2  &   $-$1.3$\pm$0.8  \\ 
UDF12-4474-6449 &   03:32:44.74 &   $-$27:46:44.9   &   7.8 &   $-$18.3 &   $-$2.9  &   $-$2.8$\pm$0.7  \\ 
UDF12-3911-6493 &   03:32:39.11 &   $-$27:46:49.3   &   7.8 &   $-$17.8 &   $-$2.8  &   $-$2.8$\pm$1.2  \\ 
UDF12-3344-6598 &   03:32:33.44 &   $-$27:46:59.8   &   7.8 &   $-$17.9 &   $-$2.7  &   $-$2.7$\pm$1.0  \\ 
UDFPar1-5644-1009 & 03:32:56.44 &   $-$27:41:00.9   &   7.8 &   $-$19.2 &   $-$2.5  &   $-$2.5$\pm$0.7  \\
UDF12-4308-6277 &   03:32:43.08 &   $-$27:46:27.7   &   8.0 &   $-$18.2 &   $-$3.6  &   $-$3.7$\pm$0.9  \\ 
UDF12-3780-6001 &   03:32:37.80 &   $-$27:46:00.1   &   8.0 &   $-$18.7 &   $-$2.4  &   $-$2.4$\pm$0.7  \\ 
UDF12-3813-5540 &   03:32:38.13 &   $-$27:45:54.0   &   8.1 &   $-$18.9 &   $-$1.4  &   $-$1.6$\pm$0.6  \\ 
UDF12-3763-6015 &   03:32:37.63 &   $-$27:46:01.5   &   8.3 &   $-$18.5 &   $-$1.9  &   $-$2.3$\pm$0.7  \\ 
\end{tabular}
\end{center}
\label{default}
\end{table*}

\end{document}